\documentstyle[prb,aps,epsf,multicol]{revtex}
\tighten
\newcommand{\beq}{\begin{equation}}
\newcommand{\eeq}{\end{equation}}
\newcommand{\bea}{\begin{eqnarray}}
\newcommand{\eea}{\end{eqnarray}}

\begin{document}
\draft

\title{
Screening of a charged particle by multivalent counterions in salty water:
Strong charge inversion}
\author
{T. T. Nguyen, A. Yu. Grosberg, and B. I. Shklovskii}
\address{Department of Physics, University of Minnesota,
116 Church
St. Southeast, Minneapolis, Minnesota 55455} \maketitle
\begin{abstract}
Screening of a macroion such as
a charged solid particle, a charged membrane, double helix DNA or
actin by multivalent counterions is considered.
Small colloidal particles, charged micelles,
short or long polyelectrolytes can play the role of
multivalent counterions.
Due to strong lateral repulsion
at the surface of macroion such multivalent counterions
form a strongly correlated liquid,
with the short range order resembling that of a Wigner crystal.
These correlations create additional binding
of multivalent counterions to the
macroion surface with binding energy larger than $k_BT$.
As a result even for a moderate concentration of multivalent
counterions in the solution, their total charge at the surface of
macroion
exceeds the bare macroion charge in absolute value. Therefore, the net
charge of the macroion inverts its
sign. In the presence of a high concentration of monovalent salt
the absolute value of inverted charge
can be larger than the bare one.
This strong inversion of charge can be observed by electrophoresis
or  by direct counting of multivalent counterions.
\end{abstract}


\begin{multicols}{2}

\section {Introduction}

Charge inversion is a phenomenon in which
a charged particle (a
macroion) strongly binds so many counterions
in a water solution that its net charge
changes sign. As shown below the binding energy of a counterion
with large charge $Z$ is larger than
$k_B T$, so that this
net charge is easily observable; for instance, it is
the net charge that determines linear transport properties, such
as particle drift in a weak field electrophoresis.
Charge inversion is possible for a variety of macroions, ranging
from the charged surface of mica or other solids
to charged lipid membranes, DNA or actin.
Multivalent metallic ions, small colloidal particles, charged micelles,
short or long polyelectrolytes can play the role of
multivalent counterions. 
Recently, charge inversion has attracted significant attention
\cite{Roland,Perel99,Shklov99,Pincus,Joanny00,Joanny1,Joanny2,Dubin,Linse}. 

Charge inversion is of special interest
for the delivery of genes to the living cell
for the purpose of the gene
therapy. The problem is that both bare DNA and a cell
surface are negatively charged and repel each other, so that DNA
does not approach the cell surface. The goal is
to screen DNA in such a way that the resulting complex
is positive~\cite{Felgner}. Multivalent counterions
can be used for this purpose. The charge inversion depends on the
surface charge density, so the cell surface charge can still
be negative when DNA charge is inverted.

Charge inversion can be also thought of as an
over-screening.  Indeed, the simplest screening atmosphere,
familiar from linear Debye-H\"{u}ckel theory, compensates at any
finite distance only a part of the macroion charge.  It can be
proven that this property holds also in non-linear
Poisson-Boltzmann (PB) theory.
The statement that the net charge
preserves sign of the bare charge agrees with the common sense.
One can think that this statement is even more universal than
results of PB equation. It was
shown~\cite{Roland,Perel99,Shklov99}, however, that this
presumption of common sense fails for screening by $Z$-valent
counterions ($Z$-ions) with large $Z$,
such as charged colloidal particles, micelles
or rigid polyelectrolytes, because there are strong
repulsive correlations between them when they
are bound to the surface of a macroion.
As a result, $Z$-ions form strongly correlated
liquid  with properties resembling a Wigner crystal (WC)
at the macroion surface.
The negative chemical potential of this
liquid leads to an additional "correlation " attraction
of $Z$-ions to the surface.
This effect is beyond the mean field PB theory,
and charge inversion is its
most spectacular manifestation.

Let us demonstrate fundamental role of lateral
correlations between $Z$-ions for a
simple model. Imagine
a hard-core sphere with radius $b$ and
with negative charge $-Q$ screened
by two spherical positive $Z$-ions with radius $a$.
One can see that if Coulomb repulsion between $Z$-ions
is much larger than $k_BT$ they are situated
on opposite sides of the negative sphere (Fig. 1a).
\begin{figure}
\epsfxsize=6.5cm \centerline{\epsfbox[85 500 565 740]{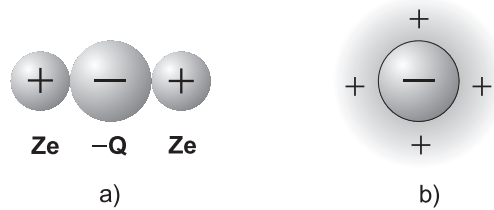}}
\caption{a) A toy model of charge inversion.
b) PB approximation does not lead to charge inversion.}
\end{figure}
If $Q > Ze/2 $, each $Z$-ion is bound because
the energy required to remove it
to infinity $QZe/(a+b) - Z^{2}e^{2}/2(a+b)$ is positive.
Thus, the charge of the whole complex $Q^* = -Q + 2Ze$
can be positive. For example, $ Q^* =3Ze/2 = 3Q$ at $Q = Ze/2$.
This example demonstrates the possibility of an
almost 300\% charge inversion. It is obviously
a result of the correlation between
$Z$-ions which avoid each other and
reside on opposite sides of the negative charge.
On the other hand, the description of screening of the central sphere
in the PB approximation smears the positive charge,
as shown on Fig. 1b and does not lead to the charge inversion. Indeed,
in this case charge accumulates in spherically symmetric screening
atmosphere only until the point of neutrality at which electric field
reverses its sign and attraction
is replaced by repulsion.

Weak charge inversion can be also obtained
as a trivial result of $Z$-ions discreteness
without correlations. Indeed, discrete $Z$-ions can
over-screen by a fraction of the "charge quantum" $Ze$.
For example, if central charge $-Q = -Ze/2$ binds one $Z$-ion,
the net charge of the complex is $Q^*=Ze/2$.
This charge is, however, three times smaller than the charge $3Ze/2$
which we obtained above for screening of the same charge $-Ze/2$ by
two correlated $Z$-ions, so that for the same $Q$ and $Z$ correlations
lead to stronger charge inversion.

Difference between charge inversion, obtained
with and without correlations becomes
dramatic
for a large sphere with a macroscopic charge $Q \gg Ze$.
In this case, discreteness by itself can lead to inverted charge
limited by $Ze$. On the other hand,
it was predicted~\cite{Shklov99} and 
confirmed by numerical simulations~\cite{Holm}
that due to correlation between $Z$-ions which
leads to their WC-like short range order
on the surface of the sphere, the net inverted charge
can reach
\begin{equation}
Q^* = 0.84\sqrt{Q Ze},
\label{Q}
\end{equation}
i. e. can be much larger than the charge quantum $Ze$. 
This charge is still smaller than $Q$ because
of limitations imposed by the very large charging energy of 
the macroscopic net charge. 

In this paper, we consider systems in which inverted charge can 
be even larger than what Eq.~(\ref{Q}) predicts.  Specifically,
we consider the problem of screening by $Z$-ions
in the presence of monovalent salt, such as NaCl, in
solution.  
This is a more practical situation than the salt-free one
considered in Ref.~\onlinecite{Perel99,Shklov99}.
Monovalent salt screens 
long range Coulomb interactions stronger than 
short range lateral correlations between adsorbed $Z$-ions. 
Therefore, screening diminishes the
charging energy of the macroion
much stronger than the correlation energy of $Z$-ions.
As a results, the inverted charge $Q^*$ becomes larger than 
that predicted by Eq. (\ref{Q}) and scales {\it linearly} with $Q$. 
The amount of charge
inversion at strong screening is limited only
by the fact that the binding energy of $Z$-ions
becomes eventually lower than
$k_BT$, in which case it is no longer meaningful 
to speak about binding
or adsorption.  Nevertheless, remaining within the strong binding
regime, we demonstrate on many examples throughout
this work, that the inverted charge, in terms of its
absolute value, can be larger than the original bare charge, sometimes
even by a factor up to 3. We call this phenomenon {\it strong} or {\it giant charge
inversion} and its prediction and theory are the main results
of our paper 
(A brief preliminary version of this paper is given in 
Ref. \onlinecite{Shklov992}).

Since, in the presence of a sufficient concentration of salt,
the macroion is 
screened at the distance smaller 
than its size, the macroion can be thought of as an
overscreened surface, with inverted charge $Q^*$ proportional to the surface
area.  In this sense, overall shape of the macroion and its surface is
irrelevant, at least to a first approximation.
Therefore, we
consider screening of a planar macroion
surface with a negative
surface charge density $-\sigma$
by finite concentration, $N$, of positive $Z$-ions,
and concentration $ZN$ of neutralizing  monovalent coions,
and a large concentration $N_1$ of a monovalent salt.
Correspondingly, we assume that all interactions are screened with
Debye-H\"{u}ckel screening length
$r\!_s = \left(8\pi l_{B}N_1\right)^{-1/2}$,
where $l_{B} = e^{2}/(Dk_B T)$ is the Bjerrum length,
$e$ is the charge of a proton, $D \simeq 80$
is the dielectric constant of water.
At small enough $r\!_s$, the method of a new boundary condition
for the PB equation suggested in Ref.~\onlinecite{Perel99,Shklov99}
becomes less convenient and in this paper
we develop more universal and direct theoretical approach
to charge inversion problem.

Our goal is to calculate the two-dimensional
concentration $n$ of $Z$-ions at the
plane as a function of $r\!_s$ and $N$.
In other words, we want to find the net charge density of the plane
\begin{equation}
\sigma^* = -\sigma + Zen.
\label{sigma*}
\end{equation}
In particular, we are interested in the maximal value of the
"inversion ratio",
$\sigma^*/\sigma$, which can be reached at large enough $N$.
The subtle physical meaning of $\sigma^{*}$ should be clearly
explained.
Indeed, the entire system, macroion plus overcharging
$Z$-ions, is, of course, neutralized by the monovalent ions.
One can ask then, what is the meaning of charge inversion?
In other words, what is the justification
of definition of Eq.~(\ref{sigma*}) which disregards
monovalent ions?

To answer we note that under realistic conditions,                               
every $Z$-ion, when on the macroion surface,                                     
is attached to the macroion with energy well in excess of $k_B T$.               
At the same time, monovalent ions, maintaining electroneutrality                 
over the distances of order $r_s$, interact with the macroion with               
energies less than $k_B T$ each. It is this very distinction that                
led us to define the net charge of the macroion including adsorbed               
$Z$-ions and excluding monovalent ions.                                          
Our definition is physically justified, it has direct experimental               
relevance.  Indeed, it is conceivable that the strongly adsorbed                 
$Z$-ions can withstand perturbation caused by the atomic force                   
microscopy (AFM) experiment, while the neutralizing atmosphere of                
monovalent ions cannot.  Therefore, one can, at least in                         
principle, count the adsorbed $Z$-ions, thus directly measuring                  
$\sigma^*$.  To give a practical example,                                        
when $Z$-ions are the DNA chains, one can realistically measure                  
the distance between neighboring DNAs adsorbed on the surface.  In               
most cases, similar logic applies to an electrophoresis experiment               
in a weak external electric field such that the current is linear                
in applied field. Sufficiently weak field does not affect the                    
strong (above $k_BT$) attachment of $Z$-ions to the macroion. In                 
other words, macroion coated with bound $Z$-ions drifts in the                   
field as a single body. On the other hand, the surrounding                       
atmosphere of monovalent ions, smeared over the distances about                  
$r_s$, drifts with respect to the macroion. Presenting linear                    
electrophoretic mobility of a macroion as a ratio of effective                   
charge to effective friction, we conclude that only $Z$-ions                     
contribute to the former, while monovalent ions contribute only to               
the latter.  In particular, and most importantly, the {\it sign}                 
of the effect - in which direction the macroion moves, along the                 
field or against the field - is determined by the net charge                     
$\sigma^{*}$ which, once again, includes $Z$-ions and does not                   
include monovalent ones. Furthermore, for a macroion with simple                 
(e.g., spherical) shape, the absolute value of the net macroion                  
charge can be also found using the mobility measurements and the                 
standard theory of friction in electrolytes \cite{Hunter}.  This logic           
fails only for the regime which we call strongly non-linear.  In                 
this regime, majority of monovalent ions form a bound Gouy-Chapman               
atmosphere of the inverted charge, and,                                          
while surface charge as counted by AFM remains equal                             
$\sigma^*$, the electrophoretic measurement yields universal                     
value $e/2\pi l_{B}r_s$, which is inverted but is smaller than 
$\sigma^*$. For a macroion of the size smaller than $r_s$, its size              
determines the maximum inverted charge.

Now, as we have formulated major goal of the paper, let
us describe briefly its structure and main results.
In Sec. II - IV
we consider screening of a charged surface by compact $Z$-ions such as
charged colloidal particles, micelles or
short polyelectrolytes, which can be modeled as a sphere with radius
$a$. We call such $Z$-ions "spherical".
Spherical ions form correlated liquid with properties similar to
two-dimensional WC (Fig. 2).
\begin{figure}
\epsfxsize=5cm \centerline{\epsfbox[150 555 300 660]{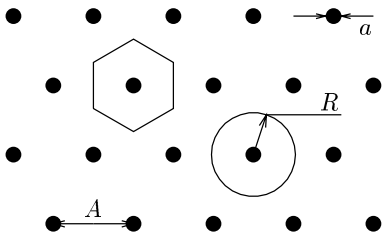}}
\caption{Wigner crystal of $Z$-ions on
the background of surface charge.
A hexagonal Wigner-Seitz cell and its simplified version as
a disk with radius $R$  are shown.}
\end{figure}
In Sec. II we begin with screening of the simplest
macroion which is a thin charged sheet
immersed in water solution (Fig. 3a).
This lets us to postpone the complication related to
image potential which appears for
a more realistic macroion which is a thick insulator
charged at the surface (Fig. 3b).
We calculate analytically the dependence of the
inversion ratio, $\sigma^{*}/\sigma$, on
$r\!_s$ in two limiting cases
$r\!_s\gg R_0$ and $r\!_s\ll R_0$,
where $R_0 = (\pi \sigma/ Ze)^{-1/2}$ is the
 radius of a Wigner-Seitz cell
at the neutral point $n=\sigma/Ze$ (we approximate
the hexagon by a disk).
We find that at $r\!_s\gg R_0$
\begin{equation}
\sigma^{*}/\sigma =  0.83 (R_0/r\!_s) = 0.83 \zeta^{1/2},~~~(\zeta \ll
1)
\label{smallyI}
\end{equation}
where $\zeta = Ze/\pi \sigma r\!_s\!^2 = (R_0/r\!_s)^{2}$.
At $r\!_s\ll R_0$
\begin{equation}
\frac{\sigma^{*}}{\sigma}
=\frac{2\pi\zeta}{\sqrt3~ \ln^{2}\zeta},~~~(\zeta\gg 1).
\label{giantI}
\end{equation}
Thus $\sigma^{*}/\sigma$ grows with decreasing
$r\!_s$ and can become larger than 100\%.
We also present numerical calculation
of the full dependence of the inversion ratio on $\zeta$.

\begin{figure}
\epsfxsize=6cm \centerline{\epsfbox[0 0 450 380]{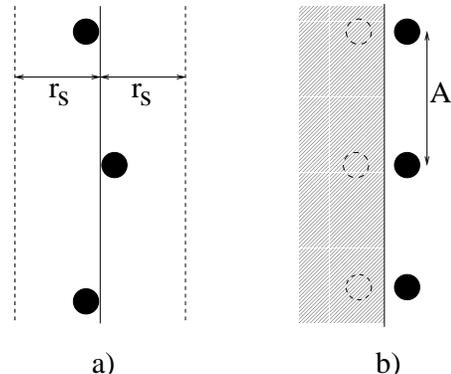}}
\caption{Two models of a macroion
studied in this paper. $Z$-ions are shown
by full circles. a) Thin charged plane
immersed in water. The dashed lines show the position
of effective capacitor plates related to the screening charges.
b) The surface of a large macroion. Image
charges are shown by broken circles.}
\end{figure}

In Sec. III we discuss effects related to finite size of $Z$-ion.
It is well known~\cite{Gueron} that monovalent ions can
condense on the surface of a small
and strongly charged spherical $Z$-ion.
As a result, instead of the bare charge of $Z$-ions
in Eqs.~(\ref{smallyI})
and (\ref{giantI}) one should use the
net charge of $Z$-ions, which is
substantially smaller. Thus, condensation puts a limit
for the inversion ratio. The net charge grows with
the radius $a$ of the $Z$-ion. Therefore, we
study in this section the case when
$r\!_s\ll a \ll R_0$ and showed that the largest inversion
ratio for spherical ions can reach a few hundred percent.

Sec. IV is devoted to more realistic macroions
which have a thick
insulating body with dielectric constant much smaller than that of
water.
In this case each $Z$-ion has an image charge of the same sign
and magnitude.
 Image charge repels $Z$-ion and pushes WC
away from the surface.  In this case charge inversion is
studied numerically in all the range of $r\!_s$
or $\zeta$. The result turns out to be
remarkably simple: at  $\zeta < 100$,
the inversion ratio is
twice smaller than for the case of the
charged sheet immersed in water. A simple
interpretation of this result will be given in Sec. IV.

In Sec. V and VI we study adsorption of long
rod-like $Z$-ions  with negative
linear charge bare density $-\eta_0$ on a surface with a
positive charge density $\sigma$.
(We changed the signs of both surface and $Z$-ion charges
to be closer to the practical case when DNA double helices are adsorbed
on a positive surface.) Due to the strong lateral
repulsion, charged rods tend to be parallel to each other and
have a short range order of an one-dimensional WC (Fig. 4).
In the Ref.\onlinecite{Yang} one can find
beautiful atomic force microscopy pictures of almost
perfect one-dimensional WC of DNA double
helices on a positive membrane.
The adsorption of another rigid polyelectrolyte, PDDA, was studied in
Ref.~\onlinecite{mashl}. Here we concentrate on the case of DNA.
\begin{figure}
\epsfxsize=6.5cm \centerline{\epsfbox[50 540 310 660]{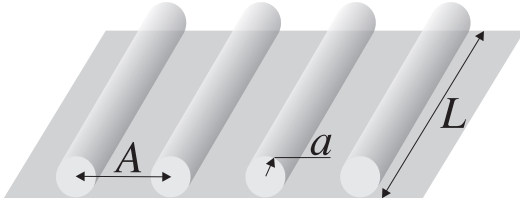}}
\caption{Rod-like negative $Z$-ions such as double helix DNA are
adsorbed on a positive uniformly charged plane.
 Strong Coulomb repulsion of rods leads to
one-dimensional crystallization with lattice constant $A$.}
\end{figure}
It is well known that for DNA, the bare charge density, $-\eta_0$
is four times larger than the critical
density $-\eta_c=-Dk_BT/e$ of the
Onsager-Manning condensation~\cite{Manning}.
According to the solution of nonlinear
PB equation,
most of the bare charge of an isolated DNA
is compensated by
positive monovalent ions residing at its surface so
that the net charge of DNA is equal to $-\eta_c$.
The net charge of DNA adsorbed on a charged surface
may differ from $-\eta_c$ due to the
repulsion
of positive monovalent ions
condensed on DNA
from the charged surface.
We, however, show
that in the case of strong screening,
$r\!_s\ll A_0$ ($A_0=\eta_c/\sigma$), the
potential of the surface is so weak that
the net charge, $-\eta$, of each adsorbed
DNA is still  equal to $-\eta_c$. Simultaneously,
at $r\!_s\ll A_0$ the Debye-H\"{u}ckel approximation
can be used to describe screening of the charged
surface by monovalent salt.
In Sec.V, these simplifications are used to study
the case of strong screening.
We show that the competition between
the attraction of DNA to the surface and the repulsion
of the neighbouring DNAs
results in the negative net surface charge density $-\sigma^*$
and the charge inversion ratio, similar to Eq.~(\ref{giantI}):
\begin{equation}
\frac{\sigma^{*}}{\sigma}=\frac {\eta_c/ \sigma r\!_s}
{\ln(\eta_c/ \sigma r\!_s)},~~~~
(\eta_c\sigma/r\!_s \gg 1)
\label{giant2}
\end{equation}
Thus the inversion ratio grows with decreasing $r\!_s$
as in the spherical $Z$-ion case.
At small enough $r\!_s$ and $\sigma$, the inversion ratio can reach
400\%. This is larger than for spherical ions because in this case,
due to the large persistence length of DNA,
the correlation energy remains large and WC-like short range
order is preserved at smaller $\sigma r_s$.
An expression similar  Eq. (\ref{giant2}) has been recently
derived for the case of polyelectrolyte with
small absolute value of the linear charge density, $\eta_0 \ll \eta_c$,
and strong screening ($r\!_s \ll A$) when screening of both
the charged surface and the polyelectrolyte
can be treated in Debye-H\"{u}ckel approximation~\cite{Joanny1}.
The result of Ref. \onlinecite{Joanny1} can be obtained
if we replace the net charge
$\eta_c$ by the bare charge $\eta_0$ in Eq. (\ref{giant2}) .

In Sec. VI we study the
adsorption of DNA rods in
the case of weak screening by monovalent salt,
$r\!_s \gg A_0$. In this case, screening of
the overcharged plane by monovalent salt
becomes strongly nonlinear, with the Gouy-Chapman
screening length $\lambda = e/(\pi l_B \sigma^*)$
much smaller than $r\!_s$. Simultaneously,
the charge of macroion repels  monovalent coions
so that some of them are released from DNA.
As a result the absolute value of the net linear charge density of
a rod, $\eta$, is larger than $\eta_c$.
We derived two nonlinear equations for unknown $\sigma^*$ and
$\eta$. Their solution at $r\!_s \gg A_0$ gives:
\begin{equation}
\frac{\sigma^{*}}{\sigma} = \frac{\eta_c}{\pi a \sigma}
\exp \left( - \sqrt {\ln\frac{r\!_s}{a} \ln\frac{A_0}{2\pi a}}\right),
\label{NL1}
\end{equation}
\begin{equation}
\eta=  \eta_c  \sqrt \frac {\ln (r\!_s/ a)}
{\ln (A_0/2\pi a)} ~~~.
\label{NL2}
\end{equation}
At $r\!_s \simeq A_0 $ we get $\eta \simeq \eta_c$, $\lambda \simeq
r\!_s$
and
$\sigma^{*}/\sigma \simeq 1$ so that Eq.~(\ref{NL1})
matches the strong screening  result of Eq.~(\ref{giant2}).
Since $\eta$ can not be smaller
than $\eta_c$, the fact that $\eta \simeq \eta_c$ already at
$r\!_s \simeq A_0$ proves that at $r\!_s \ll A_0 $,
indeed, $\eta \simeq \eta_c$

In Sec. VII we return to spherical $Z$-ions and
derive the system
of nonlinear equations which is
similar to one derived in Sec. VI
for rod-like ones. This system lets us justify
the use of Debye-H\"{u}ckel
approximation for screening of overcharged
surface ( Sec. II) at $r\!_s$ smaller than  $r_m$, where
$r_m = a\exp(R_0/1.65 a)$ is an exponentially large length.
We show that even at $r\!_s \gg r_m$
nonlinear equations lead only to a small
correction to the power of $r\!_s$ in
Eq.~(\ref{smallyI}).

In Sec. I-VII we assume that  the surface charges
of a macroion are frozen and can not move. In Sec. VIII we explore
the role of the mobility of these charges.
Surface charge can be mobile, for example, on charged
liquid membrane where hydrophilic heads can move along the surface.
If a membrane surface has heads with two different charges,
for example, 0 and -$e$, the negative ones can replace
the neutral ones near the positive $Z$-ion,
thus accumulating around it and
binding it stronger to the surface.
We show that this effect enhances
charge inversion substantially.
We conclude in Sec. IX. 

\section {Screening of charged sheet by spherical $Z$-ions}

Assume that a  plane with the charge density $-\sigma$
is immersed in water (Fig. 3a) and is
covered by $Z$-ions with two-dimensional concentration $n$.
Integrating out all the monovalent ion degrees of freedom, or,
equivalently, considering all interactions screened at the distance
$r\!_s$, we can write down the free energy per unit area in the form
\begin{equation}
{\mathcal F} = \pi \sigma^{2} r\!_s/D  -2 \pi \sigma r\!_s Zen/D + 
F_{ZZ} + F_{id},
\label{free}
\end{equation}
where the four terms are responsible, respectively, for
the self interaction of the charged plane,
for the interaction  between
$Z$-ions and the plane, for pair interactions
between $Z$-ions and for the entropy of ideal two-dimensional gas of
$Z$-ions.
Using Eq.~(\ref{sigma*}) one can rewrite Eq.~(\ref{free}) as
\begin{equation}
{\mathcal F}= \pi (\sigma^{*})^{2} r\!_s/D + F_{OCP},
\label{free1}
\end{equation}
where $F_{OCP} = F_{c}+ F_{id}$ is the free energy
 of the same system of $Z$-ions
residing on a neutralizing background with surface charge density
$-Zen$, which is conventionally referred to as
one component plasma (OCP), and
\begin{equation}
F_{c}= -\pi (Zen)^2 r\!_s/D  + F_{ZZ}
\label{free2}
\end{equation}
is the correlation part of $F_{OCP}$.
The transformation from Eq.~(\ref{free}) to Eq.~(\ref{free1})
 can be simply interpreted as the
addition of uniform charge densities $-\sigma^*$
and $\sigma^*$ to the plane.
The first addition makes a neutral OCP on the plane.
The second addition creates two planar capacitors
with negative charges on both sides of the plane which screen
the inverted charge of the plane at the distance $r\!_s$ (Fig. 3a).
The first term of Eq.~(\ref{free1})
is nothing but the energy of these two capacitors.
There is no cross term corresponding to the interactions
between the OCP and the capacitors
because each planar capacitor creates a constant potential,
$\psi(0) = 2\pi \sigma^{*} r\!_s/D$, at the neutral OCP.

Using Eq.~(\ref{free2}), the electrochemical potential of $Z$-ions
 at the plane can be written as
$\mu = Ze \psi(0) + \mu_{id}  + \mu_{c}$,
where $\mu_{id}$ and $\mu_{c}=\partial F_{c}/\partial n$ are
the ideal and the correlation parts of the chemical potential of OCP.
In equilibrium, $\mu$ is equal to the chemical
potential, $\mu_{b}$, of the ideal bulk
solution,
because in the bulk electrostatic potential $\psi=0$.
Using Eq. (\ref{free1}), we have:
\begin{equation}
2\pi \sigma^{*} r\!_s Ze/D = - \mu_{c} + (\mu_{b}- \mu_{id}).
\label{master}
\end{equation}
As we show below, in most practical cases the correlation effect
is rather strong, so that $\mu_{c}$
 is negative and $|\mu_{c}|\gg k_BT$.
Furthermore, strong correlations
imply that short range order of $Z$-ions on the surface
should be similar to that of triangular Wigner crystal (WC)
since it delivers the lowest energy to OCP. Thus one can
substitute the chemical potential of
Wigner crystal, $\mu_{WC}$, for $\mu_{c}$.
One can also write the difference of ideal parts of the
bulk and the surface chemical potentials of $Z$-ions
as
\begin{equation}
\mu_{b} - \mu_{id} = k_BT \ln(N_s / N),
\label{bulkchem}
\end{equation}
where $ N_s \sim n/a$
is the bulk concentration of $Z$-ions at the plane.
Then Eq. (\ref{master})
can be rewritten as
\begin{equation}
2\pi \sigma^{*} r\!_s Ze/D = k_BT \ln(N/N_0),
\label{master1}
\end{equation}
where $N_0 = N_s \exp(-|\mu_{WC}|/k_BT)$
is the concentration of $Z$-ions in
the solution next to the charged plane.
which plays the role of
boundary condition for $N(x)$ when $x
\rightarrow 0$ \cite{Perel99,Shklov99}.
It is clear that when $N > N_0$, the net charge density
$\sigma^{*}$ is positive, i.e. has the sign opposite
to the bare charge density $-\sigma$.
The concentration $N_0$
is very small because $|\mu_{WC}|/k_BT\gg 1$. Therefore,
it is easy to achieve charge inversion.
According to Eq.~(\ref{bulkchem}) at large enough $N$
one can neglect second term of the
right side of Eq. (\ref{master}). This gives for the maximal
inverted charge density
\begin{equation}
\sigma^{*} =  \frac{D}{2 \pi r\!_s} \frac{\left| \mu_{WC} \right|}{Z e}.
\label{capacitor}
\end{equation}
Eq. (\ref{capacitor}) has a very simple meaning:
$|\mu_{WC}|/Ze$
is the "correlation" voltage which charges two above mentioned
parallel capacitors with "distance between plates" $r\!_s$ and total
capacitance per unit area $D/(2\pi r\!_s)$.

To calculate the correlation voltage $\left| \mu_{WC} \right|
/Ze$, we start from the case of weak screening when $r\!_s$
is larger than the average distance between $Z$-ions.
In this case, screening does not
affect thermodynamic properties of
WC. The energy per $Z$-ion $\varepsilon(n)$
of such Coulomb WC at $T=0$ can be estimated as
the energy of a Wigner-Seitz cell,
because quadrupole-quadrupole
interaction
between neigbouring neutral Wigner-Seitz cells
is very small.
This gives
\beq
\varepsilon(n)= -(2-8/3\pi) Z ^{2}e^{2}/RD\simeq -1.15Z ^{2}e^{2}/RD,
\label{energyWCWZ}
\eeq
where $R=(\pi n)^{-1/2}$ is
the radius of a Wigner-Seitz cell.
A more accurate calculation~\cite{mara} gives slightly
higher energy:
\beq
\varepsilon(n) \simeq - 1.11 Z ^{2}e^{2}/RD
 = - 1.96 n^{1/2}Z^{2}e^{2}/D.
\label{energyWC}
\eeq
One can discuss the role of a finite temperature on WC  in terms of
the inverse dimensionless temperature
$\Gamma = Z^{2}e^{2}/(R D k_BT)$. We are
interested in the case of large $\Gamma$.
For example, at a typical $Zen = \sigma = 1.0~e/$nm$^{2}$ and at
room temperature, $\Gamma = 10$ for $Z=4$.
Wigner crystal melts~\cite{Gann} at  $\Gamma = 130$,
so that for $\Gamma < 130$ we deal with a
strongly correlated liquid. Numerical
calculations, however, confirm that
at $\Gamma \gg 1$ thermodynamic properties of strongly correlated
liquid are close to that of WC~\cite{Totsuji}.
Therefore, for an estimate of $\mu_{c}$
we can still write  $F_{c} = n\varepsilon(n)$
and use
\begin{equation}
\mu_{WC} = \frac{\partial\left[n\varepsilon(n)\right]}{\partial n}
= - 1.65\Gamma k_BT = -1.65 \frac {Z^{2}e^{2}}{R D}.
\label{muwc}
\end{equation}
We see now that $\mu_{WC}$
is negative and $|\mu_{WC}| \gg k_BT$,
so that Eq.~(\ref{capacitor}) is justified.
Substituting Eq.~(\ref{muwc}) into Eq.~(\ref{capacitor}),
we get $\sigma^{*} = 0.83 Ze/(\pi r\!_s R)$.
At $r\!_s \gg R$, charge density $\sigma^{*}\ll \sigma$,
and $Zen \simeq \sigma$,
one can replace $R$ by $R_0= (\sigma\pi/Ze)^{-1/2}$.
This gives
\begin{equation}
\sigma^{*}/\sigma = 0.83 \zeta^{1/2},~~~(\zeta \ll 1),
\label{smally}
\end{equation}
where $\zeta = Ze/\pi \sigma r\!_s\!^2$ is the
dimensionless charge of a $Z$-ion.
Thus, at $r\!_s \gg R$ or $\zeta \ll 1$, inverted charge
density grows with decreasing
$r\!_s$. Extrapolating to $r\!_s = 2R_0$ where
screening starts to modify
the interaction between $Z$-ions substantially,
we obtain $\sigma^{*}=0.4\sigma$.

Now we switch to the case of strong screening, $r\!_s \ll R$, or
$\zeta \gg 1$. It seems that in this case $\sigma^{*}$
should decrease with decreasing  $r\!_s$, because screening
reduces the energy of WC and leads to its melting. In fact, this
is what eventually happens. However, there is
a range of $r\!_s \ll R$ where the energy of WC is still large.
In this range, as $r\!_s$ decreases, the repulsion between
$Z$-ions becomes weaker,
 what in turn makes it easier to
pack more of them on the plane.
Therefore, $\sigma^{*}$ continues to grow with
decreasing $r\!_s$.

Although we can continue to
use the capacitor model to deal with the problem, this model
loses its physical transparency when $r\!_s \ll R$, because
there is no obvious spatial separation between the
inverted charge $\sigma^*$
and its screening atmosphere. Therefore, at $r\!_s \ll R$, we
deal directly with the original free energy (\ref{free}).
The requirement that the chemical potential
of $Z$-ion in the bulk solution equals 
that of $Z$-ions at the surface now reads
\beq
\frac{\partial F}{\partial n}
= \mu_{id}-\mu_b ~~,
\label{fullcond}
\eeq
where
\beq
F= -\frac{2\pi\sigma r\!_s Zen}{D}+F_{ZZ}
\label{fint}
\eeq
is the interaction part of the total free energy (\ref{free}) apart from
the constant self-energy term $\pi\sigma^2 r\!_s/D$.
According to Eq. (\ref{bulkchem}), at large $N$ when
\beq
\mu_b -\mu_{id}=k_BT\ln(N_s/N) \ll 2\pi\sigma r\!_sZe/D ~~,
\label{cond10}
\eeq
we can neglect
the difference in the ideal part of the free energy of $Z$-ion at
the surface and in the bulk.
Therefore, the condition of equilibrium (\ref{fullcond}) can
be reduced to the problem of minimization of 
the free energy (\ref{fint})
with respect to $n$.
This direct minimization
has a very simple meaning:
new $Z$-ions are attracted to the surface,
but $n$ saturates when the increase in the repulsion energy between
$Z$-ions compensates this gain. Since this
minimization balances
the attraction to the surface with
the repulsion between $Z$-ions, the inequality (\ref{cond10})
also guarantees that thermal fluctuations of $Z$-ions around their WC 
positions are small. Therefore, $F_{ZZ}$ can be written as
\begin{equation}
F_{ZZ}=  \sum_{\bbox{r}_i \neq 0}
 \frac{(Ze)^{2}}{Dr_i}e^{-r_i/r\!_s}~~~,
\label{ZZ}
\end{equation}
where the sum is taken over all vectors of WC lattice.
At $r\!_s\ll R$, one needs to keep only interactions with 
the 6 nearest neighbours in Eq. (\ref{ZZ}). This gives
\begin{equation}
F= - \frac{2\pi \sigma r\!_s Zen}{D}  +
3n\frac{(Ze)^{2}}{DA}\exp(-A/r\!_s),
\label{SCWC}
\end{equation}
where $A = (2/\sqrt3)^{1/2}n^{-1/2}$
is the lattice constant of this WC.
Minimizing this free energy with respect to $n$ one gets
$A\simeq r\!_s\ln\zeta$,
$R \simeq (2\pi/\sqrt3)^{1/2}r\!_s\ln\zeta$ and
\begin{equation}
\frac{\sigma^{*}}{\sigma}
=\frac{2\pi\zeta}{\sqrt3~ \ln^{2}\zeta}~,~~~(\zeta\gg 1).
\label{giant}
\end{equation}

It is clear from Eq.~(\ref{giant}) that at $r\!_s \ll R$, or
$\zeta \gg 1$ the distance $R$ decreases and
inverted charge continues
to grow with decreasing $r\!_s$. This result
could be anticipated for the toy model of Fig. 1a if the
Coulomb interaction between the spheres is
replaced by a strongly screened one.
Screening obviously affects repulsion between positive spheres
stronger than their attraction to the negative one
and, therefore, makes it possible to keep two $Z$-ions even
at $Q \ll Ze$.

Above we studied analytically
two extremes,
$r\!_s \gg R$ and $r\!_s \ll R$. In the case of arbitrary $r\!_s$
we can find $\sigma^{*}$ numerically.
Indeed, minimizing the 
free energy (\ref{fint})
with the help of Eq. (\ref{ZZ}) one gets
\begin{equation}
\frac{1}{\zeta}=\sum_{\bbox{r}_i \neq 0}
\frac{3+r_i/r\!_s}{8~r_i/r\!_s} e^{-r_i/r\!_s}~~~,
\label{exacty}
\end{equation}
where the sum over all vectors of WC lattice
can be evaluated numerically. Using Eq.~(\ref{exacty})
one can find the equilibrium concentration $n$ for
any given value of $\zeta$. The resulting ratio $\sigma^*/\sigma$
is plotted by the solid curve in Fig. 5.
\begin{figure}
\epsfxsize=5.8 cm \centerline{\epsfbox[150 500 320 680]{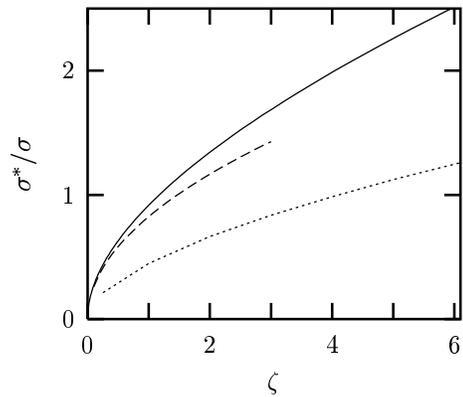}}
\caption{The ratio $\sigma^*/\sigma$ as a function
of the dimensionless charge $\zeta=Ze/\pi\sigma r\!_s\!^2$.
The solid curve is calculated for a charged plane by
a numerical solution to Eq. (\ref{exacty}),
the dashed curve
is the large $r\!_s$ limit, Eq. (\ref{smally}). The dotted curve
is calculated for the screening of the surface of the semispace
 with dielectric constant
much smaller than 80.  In this case
image charges (Fig. 3b) are taken into account (See Sec. IV).
}
\end{figure}

\section {Condensation of monovalent
coions on $Z$-ion. Role of finite size of $Z$-ion.}

We are prepared now to address the question of maximal possible
charge inversion. How far can a macroion be overcharged, and what
should one do to achieve that? We
see below that to answer this questions
one should take into account the finite size of $Z$-ions.

Fig. 5 and Eq.~(\ref{giant}) suggest
that the ratio $\sigma^*/\sigma$ continues to grow with growing
$\zeta$. However, the possibilities to increase $\zeta$ are
limited along with the assumptions of the presented theory.
Indeed, there are two ways to increase $\zeta = Z e / \pi\sigma
r\!_s\!^2$, namely to choose a surface with a small $\sigma$ or
to choose $Z$-ions with
a large $Z$. The former way is restricted because, according
to Eq. (\ref{cond10}),
$Z$-ion remains strongly bound to the charged plane only as long as
$2\pi r\!_s \sigma Z e /D \gg k_BTs$
where 
\beq
s=\ln(N_s/N)
\eeq
is the entropy loss (in units of $k_B$)
per Z-ion due to its adsorption to the surface.
This gives for $\zeta$:
\begin{equation}
\zeta \ll \zeta_{max}= 2 Z^2 l_B /s r\!_s~~.
\label{MUWC1}
\end{equation}
 Therefore, the latter way, which is to increase $Z$,
is really the most important one. The net
 charge $Z$ of a $Z$-ion is,
however, restricted because
at large charge $Z_0$ of the bare counterion
monovalent coions of the charged plane
(which have the sign opposite to $Z$-ions)
condense on the $Z$-ion surface~\cite{Gueron}.
Assuming that $Z$-ions are spheres of the
radius $a$, their net charge,
 $Z$, at large $Z_0$ can be
found from the equation
\begin{equation}
Ze^2/aD = k_BT\ln(N_{1,s} /N_1),
\label{Z}
\end{equation}
where $Ze^2/aD$ is the potential energy of a monovalent coion at the
external boundary of the condensation atmosphere ("surface") of
$Z$-ion and $k_BT\ln(N_{1,s} /N_1)$
is the difference between the chemical potentials
of monovalent coions in the bulk and at the $Z$-ion's surface,
$N_{1, s} \sim Z/a^3$  is the
concentration of coions at the surface layer.
Eq.~(\ref{Z}) gives
\begin{equation}
Z  =  \left(2a / l _B \right)\ln \left(r\!_s /a\right).
\label{Zeff}
\end{equation}
Using Eq.~(\ref{Zeff}) and Eq. (\ref{MUWC1}), we arrive at
\begin{equation}
\zeta_{max}=  \frac{8a^2}{s l_B r\!_s}
\left[ \ln \frac{r\!_s}{a} \right]^2,~~(r\!_s \gg a).
\label{MUWC4}
\end{equation}
In the theory presented in Sec. II, the radius
of  $Z$-ion, $a$, was the smallest length, even smaller than $r\!_s$.
Therefore, the largest $a$ we can substitute in Eq.~(\ref{MUWC4})
is $a =r\!_s$. For $r\!_s = a = 10 \AA$ and $s=3$
we get $\zeta_{max} \simeq 4$ so that
the inversion ratio can be as large as 2.

Since charge inversion grows with increasing $a$ we are tempted to
explore the case $r\!_s \ll a \ll R_0$.
To address this situation, our theory
needs a couple of modifications.  Specifically, in the
first term of Eq.~(\ref{SCWC}) we must take into account the fact
that only a part of a $Z$-ion interacts with the surface, namely the
segment which is within the distance $r\!_s$ from the surface.
Therefore, at $r\!_s \ll a$ results depend
on the shape of ions and distribution of charge.
If the bare charge of $Z$-ion
is uniformly distributed on the surface
of a spherical ion this adds small factor $r\!_s/2a$ to $\mu_{WC}$ and
the right side of Eq.~(\ref{MUWC1}). This gives
\begin{equation}
\zeta_{max}= Z^2 l_B /s a~~.
\label{MUWC2}
\end{equation}
One should also take into account that at $a \gg r\!_s$ Eq.~(\ref{Zeff})
should be replaced by
\begin{equation}
Z = a^2 /r\!_s l _B,
\label{Zeff2}
\end{equation}
which follows from the
condition that potential at the surface of $Z$-ion
$Ze^2/aD - Ze^2/(a+r\!_s)D$ is equal to $k_BT\ln(N_{1,s} /N_1)$.
Substituting  Eq.~(\ref{Zeff2}) to
Eq.~(\ref{MUWC2}) we find that $\zeta_{max}$
is larger than that given by Eq.~(\ref{MUWC4}), namely
\begin{equation}
\zeta_{max}=  \frac{2a^3}{s l_B r\!_s\!^2},~~(r\!_s \ll a).
\label{MUWC3}
\end{equation}
For $a = 20 \AA$, $r\!_s = 10 \AA$, $ l_B = 7 \AA$ and $s=3$
we get $\zeta_{max} \simeq 8$ so that
the inversion ratio can be as large as 3.

Let us consider now a special case of the compact $Z$-ion when it
is a {\it short} rod-like polyelectrolyte of length
$L < R$ and radius $a < r\!_s$.
Such rods lay at the surface of macroion
and form strongly correlated liquid reminding WC, so that one
can still start from Eq.~(\ref{MUWC1}).
In this case, however, Eqs. ~(\ref{Zeff}) and (\ref{Zeff2})
should be replaced by $Z \sim L \eta_c/e = L/l_B$.
Thus, $\zeta_{max} = 2 R^2/sl_B r\!_s $
and can be achieved at $L \sim R$.

We conclude this section going back to spherical $Z$-ions and
relatively weak screening.
Until now we used everywhere the Debye-H\"{u}ckel
approximation for description of
screening of surface charge density $\sigma^*$ by monovalent salt.
Now we want to verify its validity.
Theory of Sec. II requires that the correlation
voltage applied to capacitors  $|\mu_{WC}|/Ze$ is
smaller than $k_BT/e$. Using
Eqs. (\ref{capacitor}) and (\ref{muwc}) one can
rewrite this condition as $Z < R/1.65l_B$.
Substituting $Z$ from Eq.~(\ref{Zeff}) we
find that one can use linear theory only when $r\!_s < r_m$, where
\begin{equation}
r_m = a \exp(R/3.3a).
 \label{rscrit}
\end{equation}
For a large $R/2a$, the maximal
screening radius of linear theory,
$r_m$, is exponentially large.
Nonlinear theory for $r\!_s > r_m $ is given in Sec.
VII. 

\section {Screening of a thick insulating
macroion by spherical $Z$-ions: Role of images.}

In Sec. II and III we studied a charged plane
immersed in water so that screening charges are
on both sides of the plane (Fig. 3a).  In reality
charged plane is typically a surface of a rather thick membrane
whose (organic) material has the dielectric constant
$D_{1}$ much less than that of water $D_{1} \ll D$.
It is well known in electrostatics
that when a charge approaches the interface separating two
dielectrics,
it induces surface charge on interface. The
potential created  by these induced charges
can be described as the potential of an image charge sitting on the
opposite
site of the interface (Fig. 3b). At $D_{1} \ll D$,
this image charge has the same sign and magnitude as the original
charge.
Due to repulsion from images, $Z$-ions
are pushed off
the surface to some distance, $d$.
One can easily find $d$ in the case of a single
$Z$-ion near the charged macroion
in the absence of screening ($r\!_s =\infty$).
The $d$-dependent part of the free energy of this system is
\beq
F=4\pi\sigma Ze d/D + (Ze)^2/4Dd.
\label{freeone}
\eeq
Here the first term is the work needed to move $Z$-ion
from the surface to the distance $d$, and the second term is
the energy of image repulsion.
The coefficient $4\pi$ (instead of $2\pi$)
in the first term accounts for the doubling of the plane charge
due to the image of the plane.
The ion sits at distance $d=d_0$ which
minimizes the free energy of Eq. (\ref{freeone}).
Solving $\partial F/\partial d =0$,
one gets
\beq
d_0 = \frac{1}{4}\sqrt{\frac{Ze}{\pi\sigma}}=\frac{R_0}{4}.
\label{d0}
\eeq
In the presence of other counterions on the surface, the
repulsive force is stronger, therefore one expects that $d_0$ is a
little
larger than $R_0/4$.

To consider the role of all images and finite $r\!_s$,
let us start from the free energy per unit area
describing the system:
\bea
F &=& 
-\frac{4\pi\sigma r\!_sZen}{D}e^{-d/r\!_s}+
\frac{n}{2}\sum_{\bbox{r}_i \neq 0}\frac{(Ze)^2}{Dr_i}
          e^{-r_i/r\!_s} \nonumber \\
&&~~	  
+\frac{n}{2}\sum_{\bbox{r}_i}\frac{(Ze)^2}{D\sqrt{r_i^2+4d^2}}
          e^{-\sqrt{r_i^2+4d^2}/r\!_s},
\label{imageexact}
\eea
where, as in Eq. (\ref{ZZ}), the sums are taken over all vectors
of the WC lattice.
The four terms in Eq. (\ref{imageexact}) are correspondingly the self
energy of the plane, the interaction between the plane and the $Z$-ions,
the interaction between $Z$-ions (the factor $1/2$ accounts
for the double counting), and the repulsion between $Z$-ions and the
image charges (the factor $1/2$ accounts for the fact that
electric field occupies only half of the space).

At large concentration of $Z$-ions in the bulk,
the difference in the ideal parts of the free energy
of $Z$-ion in solution and at the surface can be neglected,
therefore, one can directly minimize the free
energy (\ref{imageexact}) to find the concentration of
$Z$-ions, $n$, at the surface and the optimum distance $d$.
The system of equations
\beq
\frac{\partial F}{\partial d}=0~,~~~\frac{\partial F}{\partial n} = 0,
\eeq
can be solved numerically and
the results are plotted in Fig. 5. A remarkable feature of this plot
is that, within 2\% accuracy, the ratio $\sigma^*/\sigma$
for the image problem is
equal to a half of the same ratio for the charged plane
immersed in water (for which there are no images).
If we try to interpret this result using
Eq. (\ref{capacitor}) of the capacitor model (Sec. II)
we can say that image charges do not modify the "correlation"
voltage $|\mu_{WC}|/Ze$.
The only substantial difference between two cases is that
for the thick macroion,
instead of charging two capacitors, one has to charge only
one capacitor (on one side of the surface) with
capacitance per unit area $D/4\pi r\!_s$

The fact that image charges do not modify the
"correlation voltage" can be explained quite simply in
the case of weak screening $r\!_s \gg R_0$.
In this limit, expanding the free energy (\ref{imageexact})
to the first order in $d/r\!_s$, we get
\begin{equation}
F = n \varepsilon(n)+ \frac{n}{2} Ze \phi_{WC}(n, 2d)
 +\frac{2\pi\sigma^2 d}{D}
 +\frac{2\pi(\sigma^*)^2 r\!_s}{D}.
\label{imageweak}
\end{equation}
The physical meaning of this equation is quite clear. The first two
terms are energies of the WC and of its interaction with the
image WC ($\phi_{WC}(n,2d)$ is the potential of a WC with charge
density $Zen$ at the location of an image of $Z$-ion.)
The third term is the capacitor energy created by the
charge of WC and the plane charge. And
the final term is the usual energy of a capacitor
made by the WC and the screening atmosphere.

At $\sigma^*/\sigma \ll 1$ minimization of Eq. (\ref{imageweak})
 with respect to $d$ gives the optimum
distance $d_0 = 0.3 R_0$,
which is a little larger than the estimate (\ref{d0}). Minimization
with respect to $n$ gives an equation similar to Eq. (\ref{capacitor})
\beq
\sigma^*=\frac{D}{4\pi r\!_s}\frac{|\mu_{WC}|}{Ze},
\eeq
where $\mu_{WC}$ differs from the corresponding value
in the case of immersed
plane (Eq. (\ref{muwc})) only by:
\beq
\delta \mu_{WC}=
 \frac{\partial}{\partial n}\left[
 	\frac{n}{2}Ze\psi_{WC}(n,2d_0)\right].
\eeq
It is known that $\psi_{WC}(x)$ decreases exponentially
with $x$ when $x > A/2\pi$. Since $2d_0/(A/2\pi)\simeq 1.8$, the
potential $\psi_{WC}(n,2d_0) \propto \exp(-2d_0~ 2\pi/A)$
and $\delta \mu_{WC}/|\mu_{WC}| \simeq
(1-d_0 \frac{2\pi}{A})\exp(-2d_0/(A/2\pi)) \simeq 0.02$. Thus,
at $r\!_s \gg R_0$ the chemical potential
$\mu_{WC}$ remains practically unchanged by image charges.

In the opposite limit $r\!_s \ll R_0$ one can
calculate the ratio $\sigma^*/\sigma$ by direct
minimization of the free energy, without the
use of the capacitor model.
Keeping only the nearest neighbour interactions in
Eq. (\ref{imageexact}) one finds
$$
d_0=r\!_s\ln\frac{\zeta}{8} ~~~,
$$
\beq
\frac{\sigma^*}{\sigma}\simeq\frac{2\pi\zeta}
 {\sqrt{3}\ln^2(\zeta^2/10(d/r\!_s))} \simeq\frac{\pi\zeta}
  {2\sqrt{3}\ln^2\zeta}~.
\label{imagestrong}
\eeq
Comparing this result with Eq. (\ref{giant}) for
the case of immersed plane (no image charges), one gets
\beq
\frac{(\sigma^*/\sigma)_{\it image}}{(\sigma^*/\sigma)_{\it no\,image}}
 =\frac{1}{4}\left[1+\frac{\ln 10}{\ln\zeta}\right] ~.
\label{ratio}
\eeq
Eq. (\ref{ratio}) shows that in the limit $\zeta \rightarrow \infty$,
the ratio $\sigma^*/\sigma$ for the image problem actually approaches
1/4 of that for the problem without image. However,
due to the logarithmic functions,
it approaches this limit very slowly. 
Detailed numerical calculations show that even at $\zeta=1000$,
the ratio (\ref{ratio}) is still close to 0.5.
In practice, $\zeta$ can hardly exceed $20$,
and this ratio is always close to 0.5 as Fig. 5 suggested.

Although at a given $\zeta$,
image charges do not change the results qualitatively, they,
as we show below, reduce the value of $\zeta_{max}$ substantially.
As in Sec. III, we find $\zeta_{max}$ from the condition that
the bulk electrochemical potential of $Z$-ions can be neglected.
When images are present, according to Eq. (\ref{imageexact}),
one need to replace the right hand side of Eq. (\ref{cond10})
by $2\pi\sigma r\!_s Ze \exp(-d_0/r\!_s)$.
Using Eq. (\ref{imagestrong}), this condition
now reads
\beq
\zeta \ll \zeta_{max}=4\sqrt{Z^2 l_B /sr\!_s}
\label{zetaimage}
\eeq
Using Eq. (\ref{zetaimage}) instead of Eq. (\ref{MUWC1}) and using
Eq. (\ref{Zeff}) for $Z$
we get $\zeta_{max} \simeq 5$ at
 $r\!_s = a = 10 \AA$ and $s=3$.
Therefore, according to the dotted curve of Fig. 5
which was calculated for the case of image charges,
the inversion ratio for a thick macroion can be as large as 100\%.

\section {Long charged rods as $Z$-ions.
Strong screening by monovalent salt.}

As we mentioned in Introduction
the adsorption of long rod-like $Z$-ions such as DNA double helix
on an oppositely charged surface leads to the strong charge
inversion. In this case, correlations
between rods cause parallel ordering of rods
in a strongly correlated nematic liquid.
In other words, in the
direction perpendicular to the rods we deal with short range order of
one-dimensional WC (Fig. 4).

Consider the problem of screening of a positive plane
with surface charge density, $\sigma$,
by negative DNA double helices with the net linear charge density $-\eta$
and the length $L$ smaller than
the DNA persistence length $L_p$
so that they can be considered straight
rods.
For simplicity, the charged plane
is assumed to be thin and immersed in water so that we can neglect
image charges. Modification of the results due to image
charges is given later. Here, the
strong screening case $r\!_s \ll A$ is considered
($A$ is the WC lattice constant). The
weak screening case, $r\!_s \gg A$, is the topic of the next section.

We show below that at $r\!_s \ll A$ screening radius $r\!_s$ is
smaller than the Gouy-Chapman length for the bare plane. Therefore,
one can use Debye-H\"{u}ckel formula, $\psi(0)=2\pi\sigma r\!_s/D$,
for the potential of the plane.
On the other hand, the "bare" surface charge of DNA is very large,
and its corresponding
Gouy-Chapman length is much smaller
than $r\!_s$. As the result, one
needs nonlinear theory for description of the net charge of DNA.
It leads to Onsager-Manning conclusion
 that positive monovalent ions condense on the
surface of DNA reducing its net charge, $-\eta$, to
$-\eta_c= -Dk_BT/e$. Far away from DNA, the
linear theory can be used.
When DNA rods condense on the plane, we can still use $-\eta_c$ as
the net charge density of DNA, because as we will see later, the
strongly screened potential of plane only weakly
affects condensation of monovalent ions on DNA.

Therefore, we can write the free energy per DNA as
\beq
f=-\frac{2\pi\sigma r\!_sL\eta_c}{D}   + \frac{1}{2}
\sum_{\parbox{0.39in}{\scriptsize
$i=-\infty$\\ $~~~i\neq 0$}}
^{\infty}
\frac{2L\eta_{c}^{2}}{D} K_0\left(\frac{iA}{r\!_s}\right),
\label{rodfree}
\eeq
where $K_0(x)$ is the modified Bessel function of $0$-th order.
The first term of Eq. (\ref{rodfree})
describes the interaction energy of
DNA rods with the charged plane, the second term
describes the interaction between DNA rods
arranged in one-dimensional WC,
the factor $1/2$ accounts for the double counting of
the interactions in the sum.

Since the function $K_0(x)$ exponentially decays at large $x$,
at $r\!_s \ll A$ one can keep only the
nearest neighbour interactions in
Eq. (\ref{rodfree}). This gives
\beq
f\simeq
  -\frac{2\pi\sigma r\!_sL\eta_c}{D} + \frac{2L\eta_c^2}{D}
   \sqrt{\frac{\pi r\!_s}{2A}}\exp(-A/r\!_s)~,
\label{rodA}
\eeq
which is similar to Eq. (\ref{SCWC}).
To find $A$, we minimize the free energy per unit area,
$F=nf$, with respect to $n$,
where $n=1/LA$ is the concentration of DNA helices
at the charged plane. This yields:
\beq
\frac{\sqrt{2\pi}\sigma r\!_s}{\eta_c}=\sqrt{A/r\!_s}\exp(-A/r\!_s).
\label{rodAA}
\eeq
Calculating the net negative surface charge density, $ -\sigma^*=
-\eta_c/A+\sigma$,
we obtain for the inversion ratio
\beq
\frac{\sigma^*}{\sigma} \simeq
 \frac{\eta_c/\sigma r\!_s}{\ln(\eta_c/\sigma r\!_s)}
~~~~(r\!_s\ll A).
\label{DNAstrong}
\eeq
As we see from Eq. (\ref{rodAA}), the lattice constant $A$ of WC
decreases with decreasing $r\!_s$ and charge inversion becomes stronger.

Let us now address the question of the maximal charge inversion
in the case of screening by DNA.
Similar to what was done in Sec. III, the charge inversion ratio is
limited by the condition that the electrochemical
potential of DNA in the bulk solution can be neglected and
therefore, DNA is strongly bound to the surface.
Using Eq. (\ref{rodA}) and (\ref{rodAA}), this
condition can be written by an equation similar to Eq. (\ref{cond10})
\beq
k_BT s \ll 2\pi\sigma r\!_s L\eta_c/D
 ~~$or~~$ \eta_c/\sigma r\!_s \ll 2\pi L/s l_B,
\label{cond1}
\eeq
where $s = \ln(N_{s,DNA}/N_{DNA})$ is the entropy loss
(in units of $k_B$) per DNA due
to its adsorption to the surface.
$N_{s,DNA}$ and $N_{DNA}$ are correspondingly
the three-dimensional concentration of DNA at
the charged surface and in the bulk.
Inequality (\ref{cond1}) also guarantees that
WC-like short range order of DNA helices
is preserved.
To show this, let us assume that the left and right
nearest neighbour rods at the surface are
parallel to each other and discuss the amplitude of
the thermal fluctuations of the
central DNA along the axis $x$ perpendicular to DNA direction
(in the limit $r\!_s \ll A$, we need to deal only with
two nearest neighbours of the central DNA).
At $x=0$, the free energy of the rod is given by Eq.
(\ref{rodA}). At $x \neq 0$ the free energy of the central DNA is
\beq
f(x)\simeq
  -\frac{2\pi\sigma r\!_s\eta_{c}L}{D} +
  \frac{2L\eta_c^2}{D}
   \sqrt{\frac{\pi r\!_s}{2A}} \cosh
   \left(\frac{x}{r\!_s}\right)e^{-A/r\!_s}.
\label{rodA1}
\eeq
Using Eqs. (\ref{rodA1})
and (\ref{rodAA}),
we find the average
amplitude, $x_0$, of the fluctuations of $x$ from the condition
$f(x_0)-f(0) \simeq k_BT$. This gives 
$x_0\simeq r\!_s\ln(Ae/2\pi\sigma r\!_s\!^2L)$.
The inequality (\ref{cond1}) then gives:
\beq
x_0 < r\!_s \ln\frac{A}{r\!_s} \simeq r\!_s \ln\ln
\frac{\eta_c}{\sigma r\!_s}
\ll A\simeq r\!_s \ln\frac{\eta_c}{\sigma r\!_s}.
\label{rodA3}
\eeq
Thus, DNA helices preserve WC-like short range
order when the condition (\ref{cond1}) is met.

This condition  obviously
puts only a weak restriction on maximum value
of $\sigma^*/\sigma$.
At $L = L_p = 50$~nm and $s=3$, the parameter $\eta_c/\sigma
r\!_s$ can be as large as 75 and, according to Eq. (\ref{DNAstrong})
the ratio $\sigma^*/\sigma$ can reach 15.
Therefore, we can
call this phenomenon {\it strong charge inversion}.

This limit can be easily reached at a very
small $\sigma$. On the other hand, if we want to reach it making $r\!_s$
very small we have to modify this theory for the case when
$r\!_s$ is smaller than the radius of DNA.
In a way, this is similar to what was done in Sec. 3 for spherical
$Z$-ions. At $r\!_s \ll a$ one replaces the
net charge of DNA, $\eta_c$ by $\eta_c a/r\!_s$ and adds small factor
$(r\!_s/\pi^{2}a)^{1/2}$
to the first term of Eq. (\ref{rodA}).
This modification changes only logarithmic
 term of Eq. (\ref{DNAstrong})
and does not change our conclusion about strong charge inversion.
\begin{figure}
\epsfxsize=5.8 cm \centerline{\epsfbox[150 500 320 670]{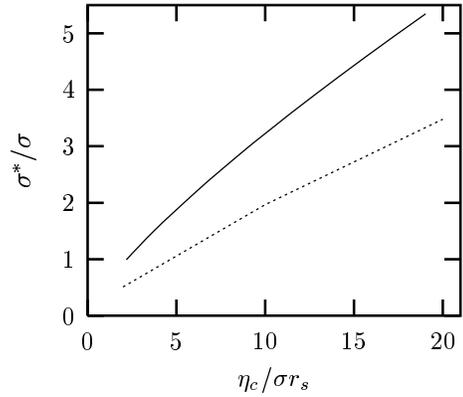}}
\caption{The ratio $\sigma^*/\sigma$ as a function
of $\eta_c/\sigma r\!_s$.
The solid curve is calculated for a charged plane by
numerical solution to Eq. (\ref{rodfree}).
The dotted curve
is calculated for the screening of the surface of the semi-space
with dielectric constant
much smaller than 80.  In this case
image charges are taken into account.
} \end{figure}

One can numerically minimize the free energy (\ref{rodfree}) at all
$r\!_s \leq A$ to find $\sigma^*/\sigma$. The result is plotted by
the solid curve in Fig. 6.

Let us now move to the more realistic case of a thick macroion,
so that repulsion from image charges must be taken
into consideration. As in the spherical $Z$-ion case,
image charges push the WC off the surface to some distance $d$.
The free energy per DNA rod can be written as
\bea
f&=&-\frac {4\pi\sigma r\!_s L\eta_c}{D} e^{-d/r\!_s}+
\frac{1}{2}
\sum_{\parbox{0.39in}{\scriptsize
$i=-\infty$\\ $~~~i\neq 0$}}
^{\infty}
 \frac{2L\eta_c^2}{D} K_0\left(\frac{iA}{r\!_s}\right)
 \nonumber \\
&&~~
+\frac{1}{2}
\sum_{i=-\infty}^{\infty}
\frac{2L\eta_c^2}{D}
K_0\left(\frac{\sqrt{(iA)^2+4d^2}}{r\!_s}\right)~,
\label{rodimageexact}
\eea
where the three terms on the right hand side are correspondingly
the interaction between the plane and the DNA, between the different
DNAs and between the DNAs and their images.

The equilibrium distance $d_0$ and $A$ can be obtained by
minimizing the free energy per unit area $F=nf$ with respect to
$d$ and $n=1/LA$:
\beq
\frac{\partial F}{\partial d}=0~,~~~\frac{\partial F}{\partial n} = 0,
\eeq
This system of equations is solved numerically. The result
for $\sigma^*/\sigma$ is plotted by the dotted curve in Fig. 6.
It is clear that in the case of DNA, at a given value of
$\eta_c/\sigma r\!_s$, image charges play even smaller role
than for spherical $Z$-ions. The ratio $\sigma^*/\sigma$
in the case of a thick macroion is
close to 70\% of $\sigma^*/\sigma$
for the charged plane immersed in water, instead of 50\%
as in Fig. 5 for spherical $Z$-ions.

However, like in the case of spherical $Z$-ions, image charges
modify the maximal possible value of $\eta_c/\sigma r\!_s$
significantly. When images are present, according to Eq.
(\ref{rodimageexact}), one need to replace in Eq. (\ref{cond1})
$2\pi\sigma r\!_s L\eta_c/D$
by  $2\pi\sigma r\!_s L\eta_c \exp(-d_0/r\!_s)$.
Therefore, the condition that
the bulk ideal chemical potential can be neglected
and, therefore, DNA is strongly bound at the surface
has the form
\beq
k_BTs \ll 2\pi\sigma r\!_s L\eta_c \exp(-d_0/r\!_s)~~.
\label{rodcond}
\eeq
Similarly to
what was done above for the problem
of charged plane immersed in water one can show that
Eq. (\ref{rodcond}) guarantees, also,
WC-like short range order of DNA helices.
In the limit  $\eta_c \gg \sigma r\!_s$,
keeping only the nearest neighbour interactions in
the free energy (\ref{rodimageexact}) and minimizing with
respect to $d$ one gets $d_0 \simeq r\!_s \ln(\eta_c/4\sigma r\!_s)$.
Substituting $d_0$ into Eq. (\ref{rodcond}) we arrive at the
final form for the condition of Eq. (\ref{cond1}):
\beq
\eta_c/\sigma r\!_s \ll \sqrt{8\pi L/sl_B},~~~~(L \leq L_p).
\label{fincond}
\eeq
It is clear that the maximal $\eta_c/\sigma r\!_s$ and  maximal
inversion ratio grow with $L$.
For $L = L_p = 50$~nm and $s=3$,
the maximal $\eta_c/\sigma r\!_s$ = 25.
Therefore, according to the dotted curve in Fig. 6,
the inversion ratio
for a thick macroion $\sigma^*/\sigma$ can reach 4.
Such inversion can still be considered as strong.

Until now we talked about relatively short DNA, $L \leq L_p$,
which can be considered as a rod.
For DNA double-helices of a larger length ($L \gg
L_p$) the maximum inversion ratio saturates at the value
obtained above at $L = L_p$. This happens because even a long DNA
can not be adsorbed at the surface if for $L = L_p$ inequality
(\ref{fincond}) is violated. (See the theory
of adsorption-desorption phase transition,
for example, in Ref.~\onlinecite{Shura}).

On the other hand, if inequality
(\ref{fincond}) holds at $L = L_p$,
i. e. at $\eta_c/\sigma r\!_s \ll \sqrt{8\pi L_p/sl_B}$,
the adsorption of a long DNA is so strong that
DNA lays flat on the charged surface.
Since repulsion between neighbouring parallel DNA is
balanced with attraction to the surface,
interactions between parallel DNA helices are so strong that
the same inequality guarantees WC-like short range order
at the length scale $L_p$, even though DNA length
is much larger than $L_p$.
One can verify this statement studying lateral fluctuations
of a DNA segment with length $L_p$ similarly to the calculation
presented above for the problem of
charged plane immersed in water (See Eqs. (\ref{rodA1}) and
(\ref{rodA3})).
Thus, our theory and the plots of Fig. 5 are applicable for a
long DNA and, therefore, for any flexible polyelectrolyte.

To conclude this section, we would like to mention
another charge inversion problem similar
to the problem we considered here.
Giant charge inversion can be also achieved if a single very long
DNA double helix screens a long and wide
positively charged cylinder with radius
greater or about the double helix DNA persistence length (Fig. 7).
\begin{figure}[h]
\epsfxsize=4.5 cm \centerline{\epsfbox{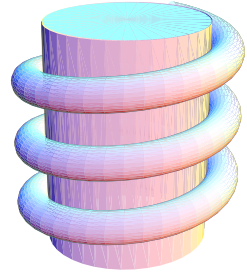}}
\caption{A long charged worm-like rods spirals around an oppositely
charged cylinder to screen it. Locally, the picture resembles
that of an one-dimensional WC}
\end{figure}
In this case, an DNA double helix spirals around the cylinder.
Neighbouring turns repel each other so that DNA forms an almost perfect
coil which locally resembles one-dimensional WC.
As a result, the cylinder charge inverts its sign:
density of DNA charge per unit length of the cylinder
becomes larger than the bare linear charge density of the cylinder.
At small $r\!_s$ this charge inversion
can be as strong as we discussed above. If cylinder diameter
is smaller than DNA persistent length
one should add elastic energy to the minimization problem.
This, of course, will make charge inversion
weaker than for wider cylinders, but still
it can be quite large.
We leave open the possibility to speculate on the relevance of these
model systems to the fact that DNA overcharges a nucleosome by
about 15\%~\cite{Pincus}.

A similar problem of
wrapping of a {\it weakly} charged polyelectrolyte
around oppositely charged sphere was recently studied
in the Debye-H\"{u}ckel approximation
in Ref.~\onlinecite{Joanny2}.
A strong charge inversion
was found in this case as well.
Charge inversion for a charged sphere
screened by an oppositely charged flexible polyelectrolyte
was previously observed in experiment~\cite{Dubin} and
numerical simulations~\cite{Linse}.

\section{Long charged rods as $Z$-ions.
Weak screening by monovalent salt}

In this section, we consider screening of a positively charged plane
by DNA rods in the case of weak screening, when $r\!_s \gg A$.
We saw in Sec. II that when the screening radius is
larger than the lattice constant of WC, the capacitor
model provides a transparent description of the charge inversion.
Here we adopt this model, too. However, we
find out that in the case of rods, the inversion charge $\sigma^*$ is
so large that its screening
by monovalent salt is nonlinear.
In other words, at $r\!_s \gg A$, the capacitors described in Sec. II
becomes nonlinear. Correspondingly in this case one has
to use the solution of the nonlinear PB equation for the plane
potential:
\beq
\psi(0)\simeq -(2k_BT/De)\ln(r\!_s/\lambda).
\label{psi0nonlinear}
\eeq
where $\lambda=e/\pi\sigma^* l_B$ is the Gouy-Chapman length.
It is shown below that
$A \ll \lambda \ll r\!_s$ so that the use of Eq. (\ref{psi0nonlinear})
is justified.

The weak screening of the plane potential
has also another important consequence.
The net charge density of DNA, $-\eta$, ceases to be equal to
to the Onsager-Manning critical density
$-\eta_c$. The charge of the plane forces DNA to release
some of monovalent coions condensed on it,
so that $\eta$ becomes larger than $\eta_c$.
Thus, in this case, we have to deal
with a nonlinear problem with two unknowns,
$\eta$ and $\sigma^*$.

One can find these unknowns from the two following
physical conditions of equilibrium.
The first one requires that
the chemical potential of positive monovalent ions (coions)
in the bulk of solution
is equal to the chemical potential of
coions condensed on the surface of DNA rods
which, in turn, are adsorbed on the plane.
The second condition requires that the chemical potentials
for DNA rods in the bulk solution
and DNA rods of the surface WC are equal.
Let us write the first condition as
\beq
k_BT \ln\frac{N_{1,s}}{N_1}= -e\psi(0)+
 \frac{2e\eta}{D}\ln(A/2\pi a),
\label{Mono_surf}
\eeq
where $N_{1}$ and $N_{1,s}$ are
the concentrations of monovalent coion
in the bulk and at the DNA surface respectively.
 The left-hand side of
Eq. (\ref{Mono_surf}) is the entropy loss
and the right-hand side is the potential energy gain
when monovalent salt condenses on the DNA surface
(the potential at the surface of DNA is the sum of
$\psi(0)$, of the nonlinear
plane capacitor made and
the potential of the DNA charged cylinder
with radius $a$ and the linear charge density $ -\eta$,
screened at the distance $A/2\pi$, by
neigbouring DNA). Far from the charged plane, DNA net charge
regains its value $-\eta_c$,
the condition of equilibrium
of condensed monovalent coions on isolated DNA rod
with those in the bulk
can be written in a way similar to Eq. (\ref{Mono_surf}):
\beq
k_BT \ln\frac{N_{1,s}}{N_1}=\frac{2e\eta_c}{D}\ln\frac{r\!_s}{a}.
\label{Mono_bulk}
\eeq
Excluding $\ln(N_{1,s}/N_1)$ from
Eqs. (\ref{Mono_surf}) and (\ref{Mono_bulk})
and using Eq. (\ref{psi0nonlinear}) we can write the first
equation for
$\lambda$ (which represents $\sigma^*$) and $\eta$ as
\beq
\eta_c \ln\frac{r\!_s}{a}=\eta_c\ln\frac{r\!_s}{\lambda}
 +\eta\ln\frac{A}{2\pi a}.
\label{first}
\eeq
The equality of the chemical potential of DNA in the bulk
and of DNA condensed on the plane can be written in the form
 similar to Eq. (\ref{capacitor})
\bea
L\eta\psi(0) &=& |\mu_{WC}|+\frac{L\eta-L\eta_c}{e}
	k_BT\ln\frac{N_{1,s}}{N_1}
	\nonumber \\
&&~ -\left(
   \frac{L\eta^2}{D}\ln\frac{\lambda}{a}
  -\frac{L\eta_c^2}{D}\ln\frac{r\!_s}{a}
 \right)~~.
\label{DNAmu}
\eea
As in Eq. (\ref{capacitor}), we see that a "correlation
voltage", $|\mu_{WC}|/L\eta$, charges
two capacitors consisting of the overcharged
 plane and its screening atmosphere to a finite
voltage $\psi(0)$. The
new second and third terms on the right hand side are due
to the change in the net charge of DNA, when it condenses on
the plane.
Specifically, the second term is the gain in the
entropy of monovalent salt released
and the third term is the loss in the self energy
of DNA when its net charge changes from $-\eta_c$ in the bulk
solution to $-\eta$ at the plane surface. Here
$\lambda$ is the screening
length near the plane surface.
(This can
be seen from the fact that
the three-dimensional concentration of monovalent salt at the surface
is of the order $N_{1,s}\sim \sigma^*/2e\lambda$ and the corresponding
screening length $r_{s,surf}=(4\pi N_{1,s} l_B)^{-1/2} \sim
(2\lambda e/\pi\sigma^* l_B)^{1/2} \sim \lambda$.)

A formal derivation of Eq. (\ref{DNAmu})
is given in the end of this section.

The free energy per DNA of the one-dimensional WC of DNA rods at the
surface can be written similarly to Eq. (\ref{rodfree})
with the screening length $r\!_s$ replaced by $\lambda$,
\bea
f&=&-\frac{2\pi(\eta/A) \lambda}{D}L\eta
        +\frac{1}{2}\sum_{i=-\infty,~i\neq 0}
        ^{\infty}
                L\eta\frac{2\eta}{D} K_0\left(\frac{iA}{\lambda}\right)
	\nonumber \\
&\simeq& -\frac{L\eta^2}{D}\ln \frac{2\pi\lambda}{A}~~.
\eea
This result can be interpreted as the interaction of DNA with its
Wigner-Seitz cell (a stripe with length $L$, width $A$
and charge density $\eta/A$).

The chemical potential $\mu_{WC}$ can be easily calculated:
\beq
\mu_{WC}=\frac{\partial [nf]}{\partial n}
 \simeq -\frac{L\eta^2}{D}\ln\frac{2\pi\lambda}{A}
 ~~,
\label{muWC_rod}
\eeq
where $n=1/LA$ is the concentration of DNA at the charged surface.

Substituting Eqs. (\ref{psi0nonlinear}), (\ref{Mono_bulk}), and
(\ref{muWC_rod})
into Eq. (\ref{DNAmu}), we arrive at the second
equation for  $\eta$ and $\lambda$
\beq
2\eta\eta_c\ln\frac{r\!_s}{\lambda}=
 -\eta_c^2\ln\frac{r\!_s}{a}-\eta^2\ln\frac{A}{2\pi a}
 +2\eta_c \eta \ln\frac{r\!_s}{a}.
\label{second}
\eeq
Solving Eqs. (\ref{first}) and (\ref{second}) together with
$A=\eta/(\sigma+\sigma^*)$, we get
\beq
\eta\simeq\eta_c\sqrt{\frac{\ln(r\!_s/a)}{\ln(A_0/2\pi a)}},
\label{DNAweak1}
\eeq
\beq
\ln\frac{\lambda}{a}\simeq\sqrt{\ln\frac{r\!_s}{a}\ln\frac{A_0}{2\pi
a}},
\label{DNAweak2}
\eeq
where $A_0=\eta_c/\sigma$.

Eq. (\ref{DNAweak2}) shows that the theory is self consistent:
when $r\!_s \gg A_0$, one has $r\!_s \gg \lambda \gg A_0$. This
justifies
the use of nonlinear potential for the plane. Eq. (\ref{DNAweak1})
demonstrates that $\eta \gg \eta_c$ as we anticipated.
Eq. (\ref{DNAweak1}), of course, is valid only
if $\eta \leq \eta_0$, where
$\eta_0$ is bare linear charge density of DNA.

The ratio $\sigma^*/\sigma$ can now be easily calculated
by substituting $\lambda=e/\pi \sigma^* l_B$ into Eq. (\ref{DNAweak2}).
One arrives at Eq. (\ref{NL1}) which shows that the
ratio $\sigma^*/\sigma$
increases as $r\!_s$ decreases, but remains smaller than unity.
 When $r\!_s \sim A_0$
one finds from Eqs. (\ref{DNAweak1}) and (\ref{DNAweak2}) that
$\eta\sim \eta_c$, $\lambda\sim r\!_s\sim A_0$,  and
$\sigma^*/\sigma \sim 1$, what
matches the Eq. (\ref{DNAstrong})
obtained for the strong screening limit ($r\!_s \ll A$).

Let us now present a derivation of Eq. (\ref{DNAmu}).
To calculate the free energy of the system
we use the standard charging procedure described,
for example, in Ref.~\onlinecite{Harned} and used for DNA in
Ref.~\onlinecite{Fixman,Gueron1}.
First, let us start by calculating the
electrostatic free energy of a DNA dissolved
in solution, which can be written as the work needed to
charge the DNA up to the bare value $\eta_0$ per unit length
\beq
f=L\int_0^{\eta_0}\phi(\eta^\prime) d\eta^\prime ~~,
\eeq
where $\phi(\eta^\prime)$ is the self consistent surface potential of
DNA
when its charge is $\eta^\prime$ per unit length.
Following Ref.~\onlinecite{Gueron1},
let us divide this charging process in two steps. First, the DNA
is charged from $0$ up to $\eta_c$. In this step, one can use for
$\phi(\eta^\prime)$
the linear (Debye-H\"{u}ckel) potential
\beq
\phi(\eta^\prime)=\frac{2\eta^\prime}{D}
\frac{K_0(a/r\!_s)}{K_1(a/r\!_s)a/r\!_s}
\simeq \frac{2\eta^\prime}{D}\ln(r\!_s/a),~~(r\!_s \gg a).
\eeq
In the next step, DNA is charged from $\eta_c$ to $\eta$. In this step,
one has to use nonlinear potential for $\phi(\eta^\prime)$.
It can be written as a sum
\beq
\phi(\eta^\prime)=2\frac{k_BT}{De}\ln\frac{a}{\Lambda(\eta^\prime)}
  + \frac{2\eta_c}{D}\ln\frac{r\!_s}{a} ~~,
\label{phinonlinear}
\eeq
where the first term is the
contribution of the interval $2a > r> a $ of the distances $r$
from the DNA axis. In this interval
potential can be approximated by that of the charged plane with
charge density $\eta^\prime/2\pi a$. It has Gouy-Chapman form with
the corresponding Gouy-Chapman length
$\Lambda(\eta^\prime)=a\eta_c/\eta^\prime <a$. The second term
in Eq. (\ref{phinonlinear}) is the
contribution
of interval $\infty > r> 2a$, where we deal with a cylinder of
radius $a$ and linear
net charge density $-\eta_c$.
Now, we can calculate the free energy
of a DNA rod (which
is also
the chemical potential of DNA in the bulk solution, apart from
an ideal part):
\bea
f&=&L\int_0^{\eta_c}\phi(\eta^\prime) d\eta^\prime+
        L\int_{\eta_c}^{\eta_0}\phi(\eta^\prime) d\eta^\prime \nonumber
\\
 &=&\frac{L\eta_c^2}{D}\ln\frac{r\!_s}{a}
        +\frac{L2\eta_c\eta_0}{D}\ln\frac{a}{\Lambda(\eta_0)}
        +\frac{L2\eta_c}{D}(\eta_0-\eta_c)\ln\frac{r\!_s}{a}.
	\nonumber \\
\label{DNAmufullb}
\eea
In the Onsager-Manning condensation theory, one can think
of the last two terms in the above expression as the
free energy of the condensation layer.

When DNA rods are adsorbed on the surface of the macroion, the role
of $\eta_c$ as a border between the linear and nonlinear charging
regime is played by the net charge $\eta$. We calculate the total free
energy
of the system by first charging the plane surface
to $\sigma$ and DNA to $\eta$ respectively, then continue
charging the DNA from $\eta$ to the final value $\eta_0$.
The first charging process leads to the standard contribution
$$
L\eta\psi(0)+\mu_{WC}+\frac{L\eta^2}{D}\ln\frac{\lambda}{a}~~,
$$
to the chemical potential
of DNA, where the three terms result from, correspondingly,
the capacitor energy of the screening atmosphere,
the correlation energy  of DNA and the self energy of DNA.
The second charging process builds up the condensation
layer around each DNA and gives a contribution
$$
\int_{\eta}^{\eta_0}\phi(\eta)d\eta=
        \frac{2\eta_c\eta_0}{D}\ln\frac{a}{\Lambda(\eta_0)}
        +\frac{2\eta_c}{D}(\eta_0-\eta)\ln\frac{r\!_s}{a}
$$
where the nonlinear potential of Eq. (\ref{phinonlinear}) was used.

The chemical potential of DNA on the charged surface is the sum of
the two above contributions:
\bea
&&L\eta\psi(0)+\mu_{WC}+\frac{L\eta^2}{D}\ln\frac{\lambda}{a}+
\frac{L2\eta_c\eta_0}{D}\ln\frac{a}{\Lambda(\eta_0)}
	\nonumber \\
&&~~~+\frac{L2\eta_c}{D}(\eta_0-\eta)\ln\frac{r\!_s}{a}.
\label{DNAmufulls}
\eea
Equating this expression to the chemical potential of
 DNA in the bulk
(Eq. (\ref{DNAmufullb}))
one gets the desired Eq. (\ref{DNAmu}).

So far, we have dealt only with the screening of charged
surface
by DNA double helices which
are highly charged polyelectrolytes. The situation
is simpler if one deals with weakly charged
polyelectrolytes whose bare charge density $\eta_0$ is much
smaller than $\eta_c$. In this case,
there is no condensation on the polyelectrolyte.
Therefore $\eta_0$ plays the role of the net charge $\eta_c$.
In the weak screening case, $r\!_s \gg \eta_0/\sigma$,
this brings about small changes in
Eq. (\ref{DNAmu}), which now reads:
\beq
L\eta_0\psi(0)=|\mu_{WC}|-\left(\frac{L\eta_0^2}{D}\ln\frac{\lambda}{a}
    -\frac{L\eta_0^2}{D}\ln\frac{r\!_s}{a}\right) ~.
\label{rodl_nl}
\eeq
Substituting Eq. (\ref{psi0nonlinear}) and (\ref{muWC_rod})
into Eq. (\ref{rodl_nl}),
and solving for $\lambda$, we get
\beq
\lambda \simeq
r\!_s\exp\left(\frac{\eta_0}{\eta_c}\ln\frac{\eta_0}{\sigma
r\!_s}\right)
~~(r\!_s \gg \eta_0/\sigma).
\eeq
Nonlinear effects are important when $\lambda \ll r\!_s$, or
when the exponent in the above expression becomes less than
$-1$. This gives the minimal $r\!_s$ at which nonlinear effects
are still important.
\beq
r_m = (\eta_0/\sigma) \exp(\eta_c/\eta_0) ~~.
\eeq
As we see, $r_m$ is exponentially large
at $\eta_c/\eta_0 \gg 1$. This makes this weak screening case
practically unimportant. At smaller, more realistic value
of $r\!_s$, one can use Debye-H\"{u}ckel linear theory to describe the
potential of the plane.
For $r\!_s < \eta_0/\sigma$, this has been done in
Ref.~\onlinecite{Joanny1}.
The result is an expression similar to Eq. (\ref{DNAstrong})
with the net charge $\eta_c$ replaced by the bare charge $\eta_0$.

\section {Nonlinear screening of a charged surface by spherical
$Z$-ions.}

Let us now return to the screening of the charged plane by spherical
$Z$-ions
in the case when screening by monovalent salt is very weak.
Our goal is to understand what happens
when screening radius is larger than
$r_m$ (see Eq. (\ref{rscrit})), so that
Debye-H\"{u}ckel approximation of Sec. II for the description of
screening of surface charge density $\sigma^*$ by monovalent salt
fails and a nonlinear description is necessary.

The nonlinearity of screening leads to two important changes
in the theory in Sec. II. First, the monovalent coions condense
on the surface of the $Z$-ion and reduce its apparent charge.
We discussed this condensation in Sec. III, but used
for the net charge of $Z$-ion the value obtained for isolated
 $Z$-ion in the bulk solution (Eq. (\ref{Z})).
In this section we call this charge $Z_c$ (this quantity plays
a similar role as $\eta_c$
in previous section) and save notation $Z$
for the net charge of $Z$-ion absorbed at
the charged surface as a part of the WC.
When positive $Z$-ions condense on the negative surface,
a fraction of monovalent negative ions,
condensed on the $Z$-ions is released. Therefore, strictly speaking,
$Z > Z_c$.
The charge $Z_c$ can be found from Eq. (\ref{Z}), which in the
revised notation reads
\beq
e~\frac{Z_c e}{aD} =k_BT \ln\frac{N_{1,s}}{N_1}.
\label{Znet0}
\eeq
Here, as in Sec. III, $N_{1,s}$ is the concentration of monovalent
negative ions at the external boundary of the condensation atmosphere
of the isolated spherical $Z$-ion.
The net charge of a $Z$-ion in WC, $Z$, can be found from the condition
of equilibrium of monovalent negative ions condensed on a
$Z$-ion of the WC and those in the bulk solution
\beq
\frac{Ze^2}{aD} -e\left(\frac{2.2~Ze}{RD}
+\psi(0)\right) = k_B T \ln\frac{N_{1,s}}{N_1}.
\eeq
The term in the parentheses is the total potential
of the plane and other adsorbed $Z$-ions at the considered $Z$-ion.
This potential is the
sum of the negative potential of WC
and the potential due to
the positive net charge $\sigma^*$ of the plane
given by Eq. (\ref{psi0nonlinear}).
Excluding  $k_B T \ln(N_{1,s}/N_1)$
from Eqs. (\ref{Znet0}) and (\ref{1st_eq}) we obtain
the first equation for two unknowns $Z$ and $\lambda$,
which is similar to Eq. (\ref{first}):
\beq
\frac{Ze^2}{aD} -\frac{2.2~Ze^2}{RD}
+ 2k_BT\ln(r\!_s/\lambda) = \frac{Z_{b}e^2}{aD}.
\label{1st_eq}
\eeq
To write the second equation
for these unknowns we start from the condition
that the chemical potentials of $Z$-ion at the charged surface
and in the bulk solution are equal.
In the close analogy with Eq. (\ref{DNAmu})  of Sec. VI, we can write
\bea
Ze~\psi(0)&=& |\mu_{WC}|+ (Z-Z_c)k_B T \ln\frac{N_{1s}}{N_1}
	\nonumber \\
&&~~ + \frac{(Z_{b}^2 -Z^2)e^2}{2aD}~~.
\eea
The second and third terms on the right-hand side account for
the fact that monovalent ions are released when
$Z$-ions condense on the plane surface
(so that their entropy is gained)
and simultaneously the self energy of the $Z$-ion is reduced.
Using Eqs. (\ref{muwc}), (\ref{psi0nonlinear}) and (\ref{Znet0})
we obtain the second equation for $Z$ and $\lambda$
\bea
2 k_BT Z \ln\frac{r\!_s}{\lambda} &=& \frac{1.65 (Ze)^2}{RD}+
        \frac{(Z-Z_c)Z_{b}e^2}{aD}	\nonumber \\
&&~~
        + \frac{(Z_{b}^2 -Z^{2})e^2}{2aD}~~.
\label{2nd_eq}
\eea
Solving Eqs. (\ref{1st_eq}) and (\ref{2nd_eq}) we get
\beq
\frac{Z}{Z_c}
    \simeq 1+\frac{0.56~a}{R}
\label{Zweak1}
\eeq
and
\beq
\lambda 
 = r\!_s \exp\left(-\frac{1.65 a}{2R}\ln\frac{N_{1s}}{N_1}\right)~~.
\label{Zweak2}
\eeq
Approximating $N_{1,s}$ as $N_{1,s} \sim Z/a^3$, we get
$$
\ln\frac{N_{1s}}{N} = 2\ln\frac{r\!_s}{a} ~~.
$$
Therefore
\beq
\lambda=r\!_s\exp\left(-\frac{1.65a}{R}\ln\frac{r\!_s}{a}\right)
~~.
\eeq
Nonlinear effects are important when $\lambda \ll r\!_s$, or
when the exponent in the above expression becomes
less than $-1$. This gives
 the minimal $r\!_s$ at which nonlinear effects
are still important
\beq
r_m =a\exp(R/1.65 a),
\label{rm}
\eeq
which matches the estimate Eq. (\ref{rscrit})
obtained from the side of the linear regime.

The ratio $\sigma^*/\sigma$ can be easily calculated from Eq.
(\ref{Zweak2})
\bea
\frac{\sigma^*}{\sigma}&=&\frac{e}{\pi\sigma l_Br\!_s}\exp\left(
        -\frac{1.65 a}{R}\ln\frac{r\!_s}{a}\right) 
	\nonumber \\
&=&\frac{e}{\pi\sigma l_Br\!_s}
        \left(\frac{r\!_s}{a}\right)^{-1.65a/R}
        \propto r\!_s\!^{-(1+1.65a/R)}.
\label{ocnonlinear}
\eea
Once again, this ratio increases as $r\!_s$ decreases~\cite{finite}.
Comparing Eq. (\ref{ocnonlinear})
to Eq. (\ref{smally}), we see that nonlinear effects change
the exponent in the dependence of $\sigma^*/\sigma$ on $r\!_s$ by
$1.65a/R \ll 1$. Taking into account the fact that
it is important only when $r\!_s$ is greater than an exponentially
large critical value $r_m$ (see Eq. (\ref{rm})), one can conclude from
this section that, in practical situation,
non-linear effects in the problem of screening of
a charged surface by spherical $Z$-ions are not
important.

\section {Screening of a macroion with a mobile surface charge.}

So far we have assumed that
the bare surface charges of the macroion
are fixed and can not move.
For solid or glassy surfaces,
colloidal particles and even rigid
polyelectrolytes, such as double helix DNA and actin,
this approximation seems to work well.
On the other hand, for charged lipid membranes it can be violated.
The membrane can have a
mixture of neutral and, for example, negatively charged hydrophilic
heads.
In a liquid membrane heads
are mobile so that negative ones can accumulate
near the  positive $Z$-ion and push
the neutral heads outside (see Fig. 8).
Since the background charges are now closer to the counterion,
one can immediately predict that the energy
of the WC is lower and charge inversion is stronger
than that for the case of an uniform distribution of
negative heads.

\begin{figure}
\epsfxsize=4.5 cm \centerline{\epsfbox{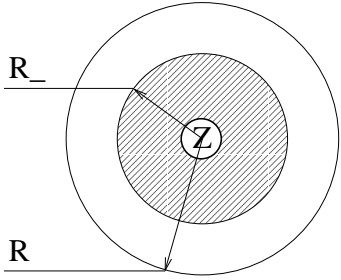}}
\vspace{0.5cm}
\caption{A $Z$-ion and its Wigner-Seitz cell with radius $R$ are shown.
The negative heads are concentrated in the shaded
area with radius $R_{-}$. The rest of the Wigner-Seitz cell
is occupied by the neutral ones.}
\end{figure}

To simplify the calculation of the free energy,
and gain more physical insight
in the problem, let us use the same transformation as in
the beginning of section II, namely we simultaneously add uniform
planar charge densities $-\sigma^*$ and
$\sigma^*$ to the plane. The first addition makes a neutral WC
on the plane. While the second addition creates the two planar
capacitors. The free energy can be written
as the sum of the energy of WC and two capacitors,
in the same way as Eq. (\ref{free1}).
Therefore, $\sigma^*$ is given by Eq. (\ref{capacitor}).

We use below the Wigner-Seitz approximation
to calculate $\mu_{WC}$. This approximation gives the
energy per ion of WC as the energy of
one Wigner-Seitz cell and neglects
the quadrupole-quadrupole
interaction between Wigner-Seitz cells.
It provides 5\% accuracy for the energy of the
standard WC on an uniform immobile background
(see Eq. (\ref{energyWC})).
In the case of mobile charges, as one sees from Fig. 7,
the quadrupole moment of the Wigner-Seitz cell is even smaller than
that for WC on an uniform background with
the same average charge density $\sigma$.
Therefore, in the case of mobile charge, the accuracy
of the Wigner-Seitz approximation is even better.

For simplicity, we assume the Wigner-Seitz cell
of a counterion is a disk with
radius $R=(\pi/n)^{1/2}$.
The negative heads concentrate around the counterion
and make a negative disk with radius $R_- < R$  and
charge density $-\sigma_-$ where
$\sigma_-=\sigma/n\pi R_-^2 \geq \sigma$. The rest of the cell
is occupied by neutral heads (Fig. 5).
The fraction of negative heads $f^2=R_-^2/R^2$ is fixed for
each membrane.
The uniform charge case is
recovered when there are no neutral heads so that $R_-=R$
and $f=1$.

Let us consider the weak screening case $r\!_s \gg R$.
Under the transformation mentioned above,
we add a disk
with radius $R$, density $-\sigma^*$ to the Wigner-Seitz cell
to neutralize it.
Now, the total energy of a Wigner-Seitz cell is the
sum of the interactions
of the $Z$-ion with two disks of radiuses $R_-$ and $R$, the self
energy of the two disks and the interaction between the disks:
\bea
\varepsilon(n)&=&-\frac{2\pi Ze\sigma_-R_-}{D}
        -\frac{2\pi Ze\sigma^*R}{D}
        +\frac{8\pi}{3}\frac{(\sigma_-)^2R_-^3}{D} 
	\nonumber \\
&&
	+\frac{8\pi}{3}\frac{(\sigma^*)^2R^3}{D}+
        \int_{(R_-)}\int_{(R)}d\bbox{r}d\bbox{r^\prime}
        \frac{\sigma_-\sigma^*}{D|\bbox{r}-\bbox{r^\prime}|}.
\label{emobile}
\eea
The integrations in the last term are taken over the disks with
radius $R_-$ and $R$ respectively. This last term can be written
as
\beq
\int_{(R_-)}\int_{(R)}d\bbox{r}d\bbox{r^\prime}
        \frac{\sigma_-\sigma^*}{D|\bbox{r}-\bbox{r^\prime}|}
= \frac{2\pi\sigma_-\sigma^*R_-^3}{D} ~{\mathcal G}(f),
\label{lastterm}
\eeq
where ${\mathcal G}(f)$ is a function of $f$ only
and can be evaluated numerically for each value of $f$ (it decreases
monotonically from $8/3$ at $f=1$ to $0$ at $f=0$).
Using $Zen=\sigma+\sigma^*$ and Eq. (\ref{lastterm}),
one gets from Eq. (\ref{emobile}):
\bea
\varepsilon(n)&=&-\frac{(Ze)^2}{RD}\left(2-\frac{8}{3\pi}\right)
        +\frac{2\pi\sigma^2R^3}{D}\left(\frac{4}{3f}+
                \frac{4}{3}-f{\mathcal G}(f)\right)
        \nonumber \\
&&      +\frac{2\pi \sigma ZeR}{D}\left(1-\frac{8}{3\pi}-\frac{1}{f}+
                \frac{f{\mathcal G}(f)}{\pi}\right).
\eea
The last two terms is the correction to $\varepsilon(n)$ due
to the mobility of the surface charge. In the uniform limit, $f=1$,
${\mathcal G}(f)=8/3$, these two terms vanish and one gets
back the usual formula for the energy per ion of WC in Wigner-Seitz
cell approximation, Eq. (\ref{energyWCWZ}).

The chemical potential for a counterion in the mobile charge
case, $\mu_{WC,m}$, can be easily calculated as
\beq
\mu_{WC,m}=\frac{\partial [n\varepsilon(n)]}{\partial n}
        \simeq -\frac{(Ze)^2}{RD}
        \left(
                2+\frac{1}{f}+\frac{4}{3\pi f}
                -\frac{2f{\mathcal G}(f)}{\pi}
        \right).
\eeq
Here $\sigma$ is approximated by $Zen$,  because
at $r\!_s \gg R$, the ratio $\sigma^*/\sigma \ll 1$.

The ratio between chemical potential $\mu_{WC,m}$
for the mobile charges
and the chemical potential $\mu_{WC}$ for the immobile
charges has been evaluated numerically as function
of the fraction $f^2$ of the negative heads. The result is plotted
in Fig. 9.
\begin{figure}
\epsfxsize=6cm \centerline{\epsfbox[100 500 300 670]{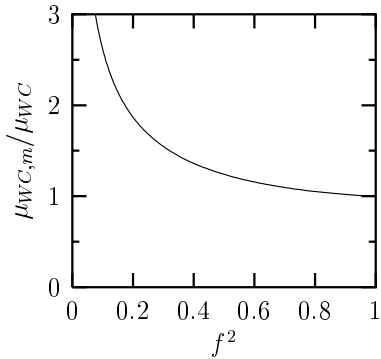}}
\caption{The ratio between the chemical potentials
$\mu_{WC,m}$ for the mobile charge
case and $\mu_{WC}$ for the immobile charge case as a
function of the fraction
of charged heads $f^2$.}
\end{figure}

Obviously, as $f$ decreases, $\mu_{WC,m}$ grows
as expected. According to Eq. (\ref{capacitor})
 this means that
the inversion ratio $\sigma^{*}/\sigma$ grows with decreasing $f$, too.
We do not continue the plot in Fig. 9 to very small $f^2$ because
in this case, the entropy of negative
heads plays important role
and screening by negative heads of the membrane can be described in
Debye-H\"{u}ckel  approximation~\cite{PincusM}.
We do not consider this regime here.

Let us now move to the limit of strong screening, $r\!_s \ll R$.
In this limit, it is more convenient
to directly minimize the free
energy, instead of using the capacitor model. Since $r\!_s \ll R$, one
needs to keep only the nearest neighbour interactions in
the free energy. Assuming $r\!_s \ll R_-$, one can write the free energy
per unit area as
\beq
F=-2\pi\sigma_-r\!_sZen + 3n \left( Ze-\frac{\sigma}{n}\right)^2
        \frac{\exp(-A/r\!_s)}{A},
\label{freemobile}
\eeq
where $Ze-\sigma/n = \sigma^*/n$ is the charge of one Wigner-Seitz cell.
In Eq. (\ref{freemobile}), the first term is the interaction of
$Z$-ion with the negative background (the disk with
charge density $\sigma_{-}$), the second term is the interaction
between neighbouring Wigner-Seitz cells.
As usual, the quadrupole-quadrupole
interaction between Wigner-Seitz cells is neglected.

Minimizing the free energy (\ref{freemobile}) with respect to $n$,
one gets $A \simeq r\!_s \ln(f^2\zeta)$ and
\beq
\frac{\sigma^{*}}{\sigma} =
\frac{2\pi\zeta}{\sqrt{3}\ln^2(f^2\zeta)}~~
,~~(\zeta \gg 1).
\eeq
where $\zeta=Ze/\pi\sigma r\!_s\!^2$.
Comparing to Eq.~(\ref{giant}), one can see that,
 as in the weak screening
case, the inversion ratio increases due to the mobility of the surface
charge.

Theory of this section is based on the assumption that
the charge of $Z$-ion is so large that it is screened nonlinearly
by the disk of opposite charge. One can easily generalize this
calculations to rod-like polyelectrolytes
and study the role of a similar stripe of positive hydrofilic heads
attracted by strongly negative DNA.
Note that the idea of nonlinear concentration of charge in membranes
with two types of heads has been used recently in a theory
of DNA-cationic lipid complexes~\cite{Bruinsma}.

\section{Conclusion}

We would like to conclude with another
general physical interpretation of the origin of charge
inversion.  To do so, let us begin with brief discussion of a separate
physical problem, namely, let us imagine that, instead of a
macroion, a neutral macroscopic metallic particle is suspended in
water with $Z$- and mono-valent ions.  In
this case, each ion creates an image charge of opposite sign
inside the metal and thus attracts to the metal.  Obviously, this
effect is by a factor $Z^2$ stronger for $Z$-ions than for
monovalent ones.  While directly at the metal surface, energy of
interaction of $Z$-ion with image, $-(Z e)^{2}/4a$, is much larger
than $k_BT$.  Therefore $Z$-ions are
strongly bound to the metallic surface, making it effectively
charged, while monovalent ions are loosely correlated with the
surface, providing for its screening over the distances of the
order of $r_s$.  We can determine the net charge of metallic
particle with bound $Z$-ions using the "capacitor model" discussed
above.  Namely, the attraction of the $Z$-ions to their images
plays the role of correlation part of the chemical potential
$\mu_c$ and provides for the voltage $Z e/4a$ which charges a
"capacitor" with the width $r_s$ between metal surface and the
bulk solution.  This leads to the result that metal surface is
charged with the net charge density $\sigma^* = Z e n = Z e/(16\pi
a r_s)$.  Note that metallic particle becomes charged due to
interactions, or correlations, between $Z$-ions and their images,
even though the particle itself was neutral in the first place.

Major results of the present paper can be now interpreted using a
similar language of images~\cite{Perel99,Shklov99}.  Although now
we consider a macroion with an insulating body, it has some bare
charge $\sigma$ on its surface, which leads to adsorption of
certain amount of $Z$-ions.  The layer of adsorbed $Z$-ions plays
the role of a metal.  Indeed, consider bringing a new $Z$-ion to
the macroion surface which has already some bound $Z$-ions. New
$Z$-ion repels nearest adsorbed ones, creating a correlation hole
for himself. In other words, it creates an image with the opposite
charge behind the surface.  Image attracts the $Z$-ion, thus
providing for the negative $\mu_c$ in Eq. (\ref{master}) and
therefore leading to the charge inversion.

The analogy between the adsorbed layer of $Z$-ions and a metal
surface holds only at length scales larger than some
characteristic length. In WC this latter scale is equal to
Wigner-Seitz cell radius $R$.
This is why for WC $\mu_c \sim -(Ze)^{2}/R$ (see Eq. (\ref{muwc})).
To make $|\mu_c|
\gg k_BT$, small enough radius $R$ is needed.  This explains why a
significant bare charge $\sigma$ is necessary to initiate
adsorbtion of $Z$-ions and to create a metallic layer with images
which can lead to charge inversion.   From formal point of view,
charge inversion in this case can be characterized by the ratio
$\sigma^{*}/\sigma$, as we did throughout the paper, while for a
neutral metallic particle such ratio is infinite.

In this paper, we considered adsorption of rigid $Z$-ions
with the shapes of either small spheres or thin rods. The concept
of effective metallic surface and image based language is
perfectly applicable in both cases.  It appears also applicable to
the other problem, not considered in this paper, namely, that of
adsorption of a flexible polyelectrolyte on an oppositely charged
dielectric macroion surface \cite{Rubinstein}.  To our mind, this
idea was already implicitly used in Ref.~\onlinecite{Joanny00},
which assumes that Coulomb self-energy of a polyelectrolyte
molecule in the adsorbed layer is negligible.  This means that
charge of the polyelectrolyte molecule is compensated by the
correlation hole, or image. It is the image charge that attracts a
flexible polyelectrolyte molecule to the surface. Interestingly,
conformations of both the polymer molecule and its image change
when the molecule approaches the surface.  

A similar role of images and correlations is actually well known
in the physics of metals. In the Thomas-Fermi approximation (which
is similar to PB one) the work function of a metal is
zero~\cite{Lang} (the work function is an analog of $\mu_c$).   The
finite value of the work function is known to result from the exchange
and correlation between electrons. For a leaving electron
it can be interpreted as interaction with its image charge in the
metal~\cite{Lang}.

We believe that interaction with image or, in other words, lateral
correlations of $Z$-ions in the adsorbed layer is the only
possible reason for a charge inversion exceeding one $Z$-ion
charge (of course, we mean here purely Coulomb systems and do not speak
about cases when charge inversion is driven by other forces, such as,
e.g., hydrophobicity).  In the Poisson-Boltzmann approximation, when
charge is
smeared uniformly along the surface, no charging of neutral metal
or overcharging of charged insulating plane is possible.

\acknowledgements

We are grateful to R. Podgornik
and I. Rouzina for useful discussions.
This work was supported by NSF DMR-9985985.

\end{multicols}

\begin{references}

\bibitem{Roland} J.\ Ennis, S.\ Marcelja and R.\ Kjellander,
Electrochim. Acta, {\bf 41}, 2115 (1996)
\bibitem{Perel99} V.\ I.\ Perel and B.\ I.\ Shklovskii, Physica A 274,
446 (1999).
\bibitem{Shklov99} B.\ I.\ Shklovskii, Phys. Rev. E{\bf 60}, 5802
(1999).
\bibitem{Pincus} E.\ M.\ Mateescu, C.\ Jeppersen and P.\ Pincus,
Europhys. Lett. {\bf 46}, 454 (1999);
S.\ Y. Park, R.\ F.\ Bruinsma, and W.\ M.\ Gelbart,
Europhys. Lett. {\bf 46}, 493 (1999);
\bibitem{Joanny00} J.\ F.\ Joanny, Europ. J. Phys. B {\bf 9} 117 (1999);
P.\ Sens, E. Gurovitch, Phys. Rev. Lett. {\bf 82}, 339 (1999).
\bibitem{Joanny1} R. R. Netz, J.\ F.\ Joanny, Macromolecules, {\bf 32},
9013 (1999).
\bibitem{Joanny2} R. R. Netz, J.\ F.\ Joanny, Macromolecules, {\bf 32},
9026 (1999).
\bibitem{Dubin} Y. Wang, K. Kimura, Q. Huang, P. L. Dubin, W. Jaeger,
Macromolecules, {\bf 32} (21), 7128 (1999).
\bibitem{Linse} T.\ Wallin, P.\ Linse, J. Phys. Chem. {\bf 100}, 17873
(1996); {\bf 101}, 5506  (1997).
\bibitem{Felgner} P.\ L.\ Felgner, Sci. American, {\bf 276},
(6) 102 (1997).
\bibitem{Holm} R.\ Messina, C. Holm, K. Kramer, Private communication.
\bibitem{Shklov992} T.\ T.\ Nguyen, A.\ Yu.\ Grosberg, B.\ I.\
Shklovskii, cond-mat/9912462.
\bibitem{Hunter} R.\ J.\ Hunter, {\it Foundations of colloid science},
Vol. 1, Oxford University Press (1986).
\bibitem{Gueron} M.\ Gueron, G.\ Weisbuch, Biopolymers,
{\bf 19}, 353 (1980); S.\ Alexander, P.\ M.\ Chaikin,
P.\ Grant, G.\ J.\ Morales, P.\ Pincus,
and D.\ Hone, J.\ Chem.\ Phys. {\bf 80}, 5776 (1984);
S.\ A. Safran, P.\ A.\ Pincus, M.\ E.\ Cates, F.\ C.\ MacKintosh,
J.\ Phys. (France) {\bf 51}, 503 (1990).
\bibitem{Yang} Ye\ Fang, Jie\ Yang, J.\ Phys.\ Chem. B {\bf 101}, 441
(1997).
\bibitem{mashl} R.\ J.\ Mashl, N.\ Gr{\o}nbech-Jensen,
M.\ R.\ Fitzsimmons, M. L\"{u}tt, DeQuan Li,
J.\ Chem. \ Phys. B {\bf 110}, 2219 (1999).
\bibitem{Manning} G.\ S.\ Manning, J.\ Chem.\ Phys. {\bf 51}, 924
(1969).
\bibitem{mara} L. Bonsall, A. A. Maradudin, Phys. Rev. B{\bf 15}, 1959
(1977).
\bibitem{Gann} R.\ C.\ Gann, S.\ Chakravarty,
and G.\ V.\ Chester, Phys.\ Rev. B {\bf 20}, 326 (1979).
\bibitem{Totsuji} H.\ Totsuji, Phys.\ Rev. A {\bf 17}, 399 (1978).
\bibitem{Shura} A. Yu. Grosberg, A. R. Khokhlov, {\it Statistical
Physics
of macromolecules}, AIP Press, New York (1994).
\bibitem{Harned} H.\ S.\ Harned, B.\ B.\ Owen, {\it Physical chemistry
of electrolytic solutions}, Reinhold, New York (1963).
\bibitem{Fixman} M.\ Fixman, J. Chem. Phys. {\bf 70}, 4995 (1979).
\bibitem{Gueron1} M.\ Gueron, J.-Ph.\ Demaret,
J. Chem. Phys. {\bf 96}, 7816 (1992).
\bibitem{finite} For a charged plane charge inversion vanishes when
$r\!_s \rightarrow \infty$. This conclusion, however can not be applied
to a finite size macroion for which excessive
charge saturates
at the value $0.84\sqrt{Q Ze}$ when $r\!_s$ exceeds the size
of the macroion~\cite{Shklov99}.
\bibitem{PincusM} R.\ Menes, N.\ Gr{\o}nbech-Jensen, P.\ Pincus,
cond-mat/9910223
\bibitem{Bruinsma} R.\ Bruinsma, Eur. Phys. J. B {\bf 4}, 75 (1998).
\bibitem{Rubinstein}M.\ Rubinstein, Private communication.
\bibitem{Lang} N.\ D.\ Lang, {\it Solid state physics}, Vol. 28, Edited by
H.\ Ehrenreich, F.\ Seitz, and D.\ Turnbull, Academic Press, New York, 1973


\end{references}
\end{document}